\documentclass[aps,prb,twocolumn,groupedaddress,showpacs,amsmath,amssymb, amsfonts,10pt]{revtex4-1}

\usepackage{adjustbox}
\usepackage[]{graphicx}
\usepackage{wrapfig}
\usepackage[caption=false]{subfig}
\usepackage{color}
\usepackage{wrapfig}
\usepackage{xcolor}
\usepackage[titletoc,toc]{appendix}
\usepackage{graphbox}

\usepackage[colorlinks,bookmarks=false,urlcolor=blue, citecolor=blue, linkcolor = blue]{hyperref}

\usepackage{tikz}
\usetikzlibrary{positioning,arrows}
\usetikzlibrary{decorations.pathmorphing}
\usetikzlibrary{decorations.markings}
\usetikzlibrary{calc}
\usetikzlibrary{patterns}
\usetikzlibrary{shapes.misc}
\tikzset{
phys/.style={thick, postaction={decorate}, decoration={markings, mark=at position .7 with {\arrow[]{triangle 45}}}},
res/.style={thick, decorate, decoration={snake,segment length=7, amplitude=1.5}},
arrow/.style={thick, draw=white, postaction={decorate}, decoration={markings, mark=at position .6 with {\arrow[black]{triangle 45}}}}
 }
 
\newcommand{\dd}{{d}}
\newcommand{\ee}{\mathrm{e}}

\newcommand{\rr}{\mathbf{r}}

\newcommand{\meff}{\widetilde{m}}
\newcommand{\geff}{\widetilde{g}}
\newcommand{\leff}{\widetilde{\lambda}}
\newcommand{\res}{\widetilde{\phi}}

\begin{document}

\title{{Universal amplitudes ratios for critical aging via functional renormalization group}\\
}
\author{Michele Vodret}
\affiliation{Chair of Econophysics and Complex Systems, \'Ecole polytechnique, 91128 Palaiseau Cedex, France.}
\author{Alessio Chiocchetta}
\affiliation{Institut f\"{u}r Theoretische Physik, Universit\"{a}t zu K\"{o}ln, D-50937 Cologne, Germany}
\author{Andrea Gambassi}
\affiliation{SISSA --- International School for Advanced Studies, via Bonomea 265, 34136 Trieste, Italy}
\affiliation{INFN, Sezione di Trieste, via Bonomea 265, 34136, Trieste, Italy.}

\begin{abstract}
We discuss how to calculate non-equilibrium universal amplitude ratios in the functional renormalization group approach, extending its applicability. In particular, we focus on the critical relaxation of the Ising model with non-conserved dynamics (model A) and calculate the universal amplitude ratio associated with the fluctuation-dissipation ratio of the order parameter, considering a critical quench from a high-temperature initial condition. Our predictions turn out to be in good agreement with previous perturbative renormalization-group calculations and Monte Carlo simulations.
\end{abstract}


\date{\today}

\maketitle

\section{Introduction}
Statistical mechanics is one of the most successful theoretical frameworks in physics, connecting the macroscopic behavior of equilibrium many-body systems to their microscopic dynamics. 
Working out this relation is, in general, challenging and it can be done either approximately or numerically at the cost of significant computational resources. 
This task is highly simplified when a system displays universality. In this case, the macroscopic, large-distance behavior does
not depend on microscopic details but rather only on some gross features of the system such as symmetries, spatial dimensionality, and range of interactions. Accordingly, different physical systems may correspond to the same universality classes, characterized by a set of critical exponents and dimensionless ratios of non-universal amplitudes~\cite{Privman1991}, or, equivalently, by scaling relations of physical quantities. 
Universal behaviors have been also observed in systems out of equilibrium, much less understood than their equilibrium counterparts. Non-equilibrium universality emerges in classical~\cite{Henkelbook1,Henkelbook2,Zia1995,Bouchaud1997,Hinrichsen2000,Tauberbook2014} and quantum~\cite{Mitra2006,Scheppach2010,Schole2012,Langen2016,Sieberer2013,Altman2015,Nicklas2015,Chiocchetta2015,Chiocchetta2017,Pruefer2018,Erne2018,Young2020,Diessel2021} systems, and even in socio-economic~\cite{Bouchaud2009} and ecologic models~\cite{ecology1,ecology2,ecology3}.  

A paradigmatic case of non-equilibrium universality is represented by the critical relaxational dynamics of statistical systems in contact with a thermal reservoir~\cite{Janssen1989,Janssen1992,Calabrese2005}. Perhaps, the simplest instance of critical dynamics is provided by the Glauber dynamics of the Ising model, characterized by the absence of conservation laws and corresponding to the so-called model A in the classification of Ref.~\onlinecite{Hohenberg1977}. 

A non-equilibrium universal behavior is displayed by this model when it is brought out of equilibrium via a critical quench, i.e., when it is prepared in an initial equilibrium state and then suddenly put in contact, at time \(t_0\), with a bath at the critical temperature. As a consequence of critical scale invariance, the equilibration time diverges, and the system undergoes slow dynamics, referred to as aging~\cite{Calabrese2005}. Aging is revealed, in the simplest instance, by two-time functions, such as the response and the correlation functions, which only depend on the ratio \(t/t'\) for $t/t' \gg 1$, with  \(t'>t_0\) the waiting time and \(t\) the observation time. 
As a consequence, the larger the waiting time \(t'\) is, the slower the system responds at time \(t\) to an external perturbation applied at time \(t'\). 
Moreover, the fluctuation-dissipation theorem, which links dynamic response and correlation functions of the system, does not apply~\cite{Tauberbook2014}, because of the breaking of time-translational symmetry (TTS) and time-reversal symmetry (TRS).  
In order to quantify the departure from equilibrium of the relaxing system, the fluctuation-dissipation ratio~(FDR) \(X\) is introduced, which differs from unity whenever the fluctuation-dissipation theorem does not apply. The long-time limit of the FDR, denoted by \(X^\infty\), has been shown to be a universal amplitude ratio, whose value does not depend on the details of the system, but only on its universality class~\cite{Calabrese2005} and possibly on some gross features of the quench~\cite{Calabrese_2006_intro,Calabrese_2007_intro}. 

Since exact solutions of the dynamics are available only for a few models, the predictions for \(X^\infty\) were obtained predominantly either via a perturbative renormalization-group (pRG) analysis or Monte Carlo (MC) simulations~\cite{Calabrese2005}. Moreover, contrary to critical exponents, the determination of amplitude ratios requires the calculation of the full form of two-time functions and, thus, it is computationally more demanding~\cite{Calabrese2002}. 

In this work, we present an approach to calculate \(X^\infty\) for the critical dynamics of model A, by using a functional renormalization-group (fRG) approach.
The application of the fRG for studying the critical short-time dynamics of model A was introduced in Ref.~\onlinecite{Chiocchetta2016},  where it was used to calculate the critical initial-slip exponent. Here, instead, we address the calculation of the universal amplitude ratio \(X^\infty\) by analyzing the long-time limit of the dynamics.
The main result of our analysis is that, within the local potential approximation (LPA)~\cite{DUPUIS20211}, the universal amplitude ratio \(X^\infty\)  depends only on the critical initial-slip exponent, leading to predictions in good agreement with the available pRG and MC estimates. While the FDR has already been studied within the fRG approach for the stationary state of the KPZ equation~\cite{Kloss2012}, we present here the first computation of \(X^\infty\) within the fRG technique for a regime where both TTS and TRS are absent.

The paper is organized as follows:  the model A, the aging dynamics, and the universality of  \(X^\infty\) are reviewed in Sec.~\ref{sec:modelA}.  The fRG approach to the study of a system undergoing a quench, as well as its LPA approximation, is discussed in Sec.~\ref{sec: funcmeth}. The calculation of the two-time functions and the connection to the aging features are discussed in Sec.~\ref{sec:5}, while our predictions for \(X^\infty\) are presented in Sec.~\ref{sec:results}. Finally, in Sec.~\ref{sec: conclusions} we summarize the approach  introduced here and we discuss future perspectives. All the relevant details of the calculations are provided in a number of appendices.

\section{Aging and  fluctuation-dissipation ratio for model A}  
\label{sec:modelA}

\subsection{Model A of critical dynamics}
Model A of critical dynamics~\cite{Hohenberg1977,Tauberbook2014} captures the universal features of relaxational dynamics in the absence of conserved quantities of a system belonging to the Ising universality class and coupled to a thermal bath. 
The effective dynamics of the coarse-grained order parameter (i.e., the local magnetization), described by the classical field \(\varphi=\varphi(\bold{r},t)\), is given by the  Langevin equation 
\begin{equation}
\label{eq:langevin}
\dot{\varphi} = - \Omega\frac{\delta \mathcal{H}}{\delta\varphi} + \zeta;
\end{equation}
here \(\dot{\varphi}\) is the time derivative of \(\varphi\), $\Omega$ is a  kinetic coefficient, $\zeta$ is a zero-mean Markovian and Gaussian noise with correlation $\langle \zeta(\rr,t)\zeta(\rr',t') \rangle = D \  \delta^{(d)}(\rr-\rr')\delta(t-t')$ and \(D\) a constant quantifying the thermal fluctuations induced by the bath at temperature $T$ (measured in units of Boltzmann constant). As long as one is interested in studying the system in the vicinity of its critical point $\mathcal{H}$ is assumed to be of the Landau-Ginzburg form: 
\begin{equation}
\label{eq:Hamiltonian}
\mathcal{H} = \int_\rr \left[ \frac{1}{2}(\nabla \varphi)^2 + \frac{\tau}{2}\varphi^2 + \frac{g}{4!}\varphi^4\right ],
\end{equation}
where $\int_\rr \equiv \int \dd^d r$ with $d$ the spatial dimensionality, $\tau$ parametrizes the distance from the critical point and $g\geq 0$ controls the strength of the interaction.  The noise \(\zeta\) is assumed to satisfy the detailed balance condition \(D=2\Omega T\)~\cite{Calabrese2005}: accordingly, the equilibrium state is characterized by a probability distribution \(\sim e^{-\mathcal{H}[\varphi]/T}\).

We assume that the system is prepared at \(t=t_0\) in an equilibrium high-temperature state with temperature \(T_0\). Accordingly, the initial value \(\varphi_0 = \varphi(\rr,t_0)\) of the order parameter can be described by a Gaussian distribution \(\sim e^{-\mathcal{H}_0[\varphi_0]/T_0}\)  with
\begin{equation}\label{eq:initial-probability}
\mathcal{H}_0 \equiv \frac{\tau_0}{2} \int_\rr \varphi^2_0.
\end{equation}

In order to study the relaxation of the system, it is convenient to consider two-time functions~\cite{Calabrese2004}, which are the simplest quantities retaining non-trivial information about the dynamics. In this work, we focus in particular on response and correlation functions of the order parameter \(\varphi\). The response function is defined as the response of the order parameter to an  external magnetic field \(h\) applied at \(\bold{r}=0\) after a waiting time \(t'>t_0\), i.e.,  
\begin{equation}
\label{defR}
\mathcal{R}_{\bold{r}}(t,t')=\frac{\delta \langle \varphi_\bold{r}(t)\rangle_h}{\delta h_\bold{0}(t')}\biggr|_{h=0},
\end{equation} 
where \(\langle \cdot \rangle_h\) stands for the mean over the stochastic dynamics induced by \(\mathcal{H}[\varphi;h]=\mathcal{H}[\phi]-  \beta \int_\rr  h\, \varphi\). The response function vanishes for \(t<t'\) because of causality. In order to simplify the notation, in what follows, we absorb the factor $\beta$ into the definition of the response function. The correlation function, instead, is defined as 
\begin{equation}
\label{defC}
\mathcal{C}_{\bold{r}}(t,t')= \langle \varphi_\bold{r}(t)\varphi_\bold{0}(t')\rangle.
\end{equation} 
The symbol \(\langle \cdot \rangle \) in Eqs.~\eqref{defR} and~\eqref{defC} indicates the  average over both the initial condition \(\varphi_0(\rr)\) and the realizations of the noise \(\zeta\). Furthermore, in Eqs.~\eqref{defR} and~\eqref{defC} we have taken advantage of spatial translational  invariance, by setting one of the spatial coordinates to zero.

 In order to characterize the distance from equilibrium of a system that is evolving in a bath at fixed temperature \(T\), the fluctuation-dissipation ratio is usually introduced~\cite{cugliandolo1994off,Calabrese2005}: 
\begin{equation}\label{chir}
X_{\rr}(t,t') \equiv \frac{ T \mathcal{R}_{\rr}(t,t')}{\partial_{t'} \mathcal{C}_{\rr}(t,t')}.
\end{equation}
When the waiting time \(t'\) is larger than the equilibration time \(t_{\text{eq}}(T)\) the dynamics is TTS and TRS. Thus, the fluctuation-dissipation theorem holds and it implies  \(X_{\bold{r}}(t,t')=1\).  The asymptotic value of the FDR, 
\begin{equation}\label{chiinfty}
X^\infty \equiv \lim_{t'\rightarrow\infty}\lim_{t\rightarrow\infty}X_\rr(t,t'),
\end{equation}
is a very useful quantity in the description of systems with slow dynamics, since \(X^\infty=1\) whenever TTS and TRS are recovered. Conversely, \(X^\infty \neq 1\) is a signal of an asymptotic non-equilibrium dynamics.
Accordingly, we distinguish such cases  as a function of the temperature \(T\) of the bath:  for  \(T>T_c\), then \(X^\infty=1\) since \(t_\text{eq}\) is finite; for \(T<T_c\), on the basis of general scaling arguments, it has been argued that \(X^\infty\) vanishes~\cite{Godreche2000}. For the special case of \(T=T_c\), there are no general arguments constraining the value of \(X^\infty\) and therefore this quantity has to be determined for each specific model.  However, \(X^\infty\) is a universal quantity  associated with critical dynamics and, in addition, the following identity holds for systems quenched to their  critical point~\cite{Calabrese2005}:  
\begin{equation}\label{chiinf}
X^\infty=\mathcal{X}^\infty_{\bold{q=0}},
\end{equation} 
where the quantity \(\mathcal{X}_\bold{q}\) is defined as in Eq.~\eqref{chir} but replacing \(\mathcal{R}_\bold{r}\) and \(\mathcal{C}_\bold{r}\) with their Fourier transforms, i.e.,  \(\mathcal{R}_\bold{q}\) and \(\mathcal{C}_\bold{q}\), respectively. Its long-time limit  \(\mathcal{X}_\bold{q}^\infty\) is obtained as in Eq.~\eqref{chiinfty}.

If the temperature  \(T\) of the bath is equal to the critical temperature \(T_c\), the equilibration time \(t_\text{eq}\) diverges and the relaxational dynamics prescribed by model A  exhibits self-similar properties, signaled by the emergence of algebraic singularities and scaling behaviors.
The scaling behavior is displayed by the response and correlation functions in two regimes:  first, the  short-time one in which \(t'\rightarrow t_m\) with fixed \(t\), where \(t_m\) is a microscopic time which depends on the specific details of the underlying microscopic model. Second, the long-time regime, in which \(t\rightarrow\infty\) with fixed \(t'\). The  response and correlation functions for model A  in the aging regime are thus given, respectively, by~\cite{Calabrese2005}
\begin{subequations}
\label{eq:greens-scaling}
\begin{align}
\mathcal{R}_{\bold{q}=0}( t, t' ) & =  \ A_\mathcal{R} (t-t')^a \left( \frac{t}{t'} \right)^\theta\, \mathcal{F}_\mathcal{R}(t'/t), \label{eq:GR-scaling}\\
\mathcal{C}_{\bold{q}=0}( t, t') & =  A_\mathcal{C} t'(t-t')^a \left( \frac{t}{t'} \right)^{\theta}\mathcal{F}_\mathcal{C}(t'/t),
 \label{eq:GK-scaling}
\end{align}
\end{subequations}
where \(A_\mathcal{R}\) and \(A_\mathcal{C}\) are non-universal amplitudes. The exponent \(a=(2-\eta-z)/z\) is associated with the time-translational invariant part of the scaling functions with \(\eta\) the anomalous dimension of the order parameter \(\varphi\), \(z\) the dynamical critical exponent while \(\mathcal{F}_{\mathcal{R}/\mathcal{C}}(t'/t)\) are universal scaling functions which satisfy \(\mathcal{F}_{\mathcal{R}/\mathcal{C}}(0)=1\). 
The breaking of TTS and TRS in the scaling form \eqref{eq:greens-scaling} is characterized by the so-called initial-slip exponent \(\theta\), which is generically independent of the static critical exponent \(\eta, \ \nu\) and of the dynamical critical exponent \(z\). 
Using the known scaling forms (\ref{eq:greens-scaling}) for the two-point functions and the relation given by Eq.~\eqref{chiinf},  one finds the following expression for the asymptotic value of the FDR \eqref{chiinf}~\cite{Calabrese2005}: 
\begin{equation}\label{Xampliratio}
X^\infty= \frac{A_\mathcal{R}}{A_\mathcal{C}(1-\theta)}.
\end{equation}
Accordingly, \(X^\infty\) is a {universal amplitude ratio}.

In Sec.~\ref{sec: funcmeth} we briefly review  the approach of Ref.~\onlinecite{Chiocchetta2016}, for studying the  critical dynamics based on the fRG technique, leaving to Sec.~\ref{sec:5} the calculation of the two-time functions (\ref{defR}) and  (\ref{defC}) in the aging regime, in which they are given by Eqs.~\eqref{eq:greens-scaling}, that finally allows us to calculate  \(X^\infty\) via Eq.~\eqref{Xampliratio}.

\subsection{Gaussian approximation}
\label{sec:gaussian}
In the absence of interaction (\(g=0\)), Eq.~\eqref{eq:langevin} is linear and therefore, by using Eqs.~\eqref{eq:initial-probability}, \eqref{defR}, and~\eqref{defC}, it is possible to calculate exactly the  response and the correlation functions. Here and in what follows the coefficient \(\Omega\) has been absorbed into the definition of the effective Hamiltonian \(\mathcal{H}\). After a Fourier transform in space with momentum \(\bold{q}\), defining \(\omega_q = q^2+r\), one finds 
\begin{align}
\label{eq:Rgaussian}
& \mathcal{R}_\bold{q}(t,t')  = \mathcal{R}_\bold{q}(t-t') = \vartheta(t-t')   e^{-\omega_q (t-t')},\\
\label{eq:Cgaussian} 
& \mathcal{C}_\bold{q}(t,t') =\mathcal{C}_\bold{q}^D(t,t')  + \tau_0^{-1} e^{-\omega_q (t+t'-2t_0)},\nonumber
\end{align}
where \(\vartheta(t)\) is the Heaviside step function and  \(\mathcal{C}^D\) is the Dirichlet correlation function 
\begin{equation}\label{Dirichlet}
\mathcal{C}_\bold{q}^D(t,t')  = \frac{1} {\omega_q} \Big[ e^{-\omega_q |t-t'|}-e^{-\omega_q (t+t'-2t_0)}\Big],
\end{equation}
which corresponds to a high-temperature (\(\tau_0 =+\infty\)) system that instantaneously looses the correlation with the initial state,  i.e., 
\begin{equation}
\mathcal{C}^D_\bold{q}(t,t_0)=0.
\end{equation}
Within the Gaussian approximation, TTS is broken by the correlation function but not by the response function. 
In addition, the dynamics~\eqref{eq:langevin} becomes critical for \(r=0\): in this case, comparing Eqs.~\eqref{eq:Rgaussian} with the scaling functions~\eqref{eq:greens-scaling}, one finds \(A_\mathcal{R}=1\), \(A_\mathcal{C} = 2\) and \(\theta=0\), which imply  
\(X^\infty =1/2\) from Eq.~\eqref{Xampliratio}.

\subsection{Response functional}
As a result of a finite interaction strength \((g\neq 0\)), the Gaussian value of the universal quantities like \(\theta\) and \(X^\infty\) may acquire sizable corrections~\cite{Calabrese2005} and a dependence on the spatial dimensionality \(d\) of the model. 
An analytical approach to treat interactions (\(g\neq 0\)) in Eq.~\eqref{eq:langevin} is provided by field-theoretical methods. From the Langevin equation~\eqref{eq:langevin} with initial condition~\eqref{eq:initial-probability} it is possible to construct the {response} action \(S=S[\tilde\varphi,\varphi]\)  given by \cite{Tauberbook2014}
\begin{align}
\label{eq:microscopic-action}
S = \mathcal{H}_0[\tilde\varphi_0,\varphi_0]  + \int_\rr \int_{t_0}^{+\infty} \widetilde{\varphi} \left(\dot{\varphi} + \frac{\delta \mathcal{H}}{\delta \varphi} - D \widetilde{\varphi}\right).
\end{align}
The scalar field  \(\tilde\varphi=\tilde\varphi(\bold{r},t)\) is the so-called response field and it appears in the definition of the response function~\eqref{defR} as:
\begin{equation}\label{Rresp}
\mathcal{R}_\bold{r}(t,t') = \langle  \varphi_\bold{r}(t) \tilde \varphi_\bold{0} (t') \rangle,
\end{equation}
where the symbol \(\langle 	\cdot \rangle\) denotes a functional integral, which,  for a generic observable \(\mathcal{O}[\varphi,\widetilde\varphi]\), can be evaluated  as 
\begin{equation}
\langle \mathcal{O}\rangle = \int \mathcal{D}\varphi \mathcal{D} \widetilde \varphi \  \mathcal{O}[\varphi,\widetilde \varphi] \  e^{-S[\varphi,\widetilde\varphi]}.
\end{equation}
This kind of average cannot, in general, be exactly computed in the  presence of interactions, and one has to resort to approximations. The action $S$ is the most suitable quantity to carry out pRG calculations of the response and the correlation function. However, for the purpose of using  fRG  it is convenient to introduce a generating function associated with the response functional \(S\), i.e., the effective action \(\Gamma\)~\cite{Tauberbook2014}.
For future convenience we denote by \(\Psi\) the vector having \(\varphi\) and \(\tilde \varphi\) as components, with \(\Psi^t = (\varphi,\tilde\varphi)\) and then indicate  \(S[\varphi,\tilde\varphi]\) as \(S[\Psi]\).

\section{Functional renormalization group for a  quench}
\label{sec: funcmeth}
In this section we introduce the fRG approach and the local potential approximation (LPA) of it, emphasizing its physical interpretation and how it can be applied to a temperature quench. The calculation of the two-time functions, instead, is presented in Sec.~\ref{sec:5}.

\subsection{fRG equation and LPA}
\label{subsec:fRG}
The fRG implements Wilson's idea of momentum shells integration at the level of the effective action \(\Gamma\)~\cite{Tauberbook2014}, for which it provides an exact flow equation. 
In order to implement the fRG scheme~\cite{Berges2002,Delamotte2007}, it is necessary to supplement the response functional $S[\Psi]$ with a cutoff function $R_k(q^2)$ over momenta \(\bold{q}\),  which  is introduced as a quadratic term in the modified action $S_k[\Psi] \equiv S[\Psi] + \Delta S_k[\Psi]$, where $\Delta S_k[\Psi] =\int_{t,\rr} \Psi^\text{t} \sigma \Psi R_k/2$ with the matrix $\sigma =\left(\begin{smallmatrix} 0 & 1 \\1 & 0\end{smallmatrix}\right)$ acting in the two dimensional space of the variable  $\Psi$.
The cutoff function $R_k$ as a function of $k$ is characterized by the following limiting behaviors~\cite{Berges2002,Delamotte2007}:
\begin{equation}\label{eq:cutoff}
\ \ R_k(q^2) \simeq
\begin{cases}
  \Lambda^2 & \text{for} \quad k\to\Lambda, \\
0 & \text{for} \quad k \to 0, \\ 
\end{cases}
\end{equation}
where $\Lambda$ is the ultraviolet cutoff, which can be identified, for instance, with the inverse of the lattice spacing of an underlying microscopic lattice model. 
The effect of $R_k$ is to supplement long-distance modes with an effective $k$-dependent quadratic term, and thus allowing a smooth approach to the critical point when the response action \(S\) is recovered for $k\to 0$. In fact, this finite $k$-dependent quadratic term regularizes the infrared divergences which would arise from loop corrections evaluated at criticality~\cite{Berges2002,Delamotte2007}. 

The running effective action $\Gamma_k$, defined from \(S_k\) as explained in App.~\ref{effective action}, can be  interpreted as a modified effective action which interpolates between the microscopic one \(S\), given here by Eq.~\eqref{eq:microscopic-action},  and the long-distance effective one \(\Gamma\); accordingly, \(\Gamma_k\) has  the following limiting behavior:
\begin{equation}\label{eq:boundaryflow}
\Gamma_k \simeq \begin{cases}
S \quad \text{for} \quad k\rightarrow \Lambda, \\
\Gamma \quad \text{for} \quad k\rightarrow 0.
\end{cases}
\end{equation}
The running effective action $\Gamma_k$ obeys the following equation~\cite{Berges2002,Delamotte2007} 
\begin{equation}
\label{eq:wetterich}
\frac{\dd \Gamma_k}{\dd k} = \frac{1}{2} \int_x \vartheta(t-t_0) \text{tr}\left(G_k \sigma \frac{\dd R_k}{\dd k}\right),
\end{equation}
where \(\int_x = \int_{\bold{r}} \int_{-\infty}^{+\infty} \dd t \), \(x=(\bold{r},t)\), and 
\begin{equation}
\label{eq:Gamma-properties}
G_k^{-1} = \Gamma^{(2)} + \sigma R_k.
\end{equation}
In Eq.~\eqref{eq:Gamma-properties} and in what follows a superscript \({(n)}\) to \(\Gamma\) denotes the \(n\)-derivative with respect to the two-dimensional variable \(\Phi = (\tilde \phi, \phi)\).
Notice that Heaviside theta in Eq.~\eqref{eq:wetterich} enforces the quench protocol described in Sec.~\ref{sec:modelA}~\cite{Chiocchetta2016}.
Equation ~\eqref{eq:wetterich} is a (functional) first-order differential equation in $k$. Accordingly, in order to be solved for \(\Gamma_{k}\), one has to specify an initial condition for the RG flow, i.e.,  one has to set the coupling constants which appear in  Eq.~\eqref{eq:Hamiltonian} at scale \(k=\Lambda\) in the corresponding effective action \(\Gamma_{\Lambda}=S\) (see Eq.~\eqref{eq:boundaryflow}).

While Eq.~\eqref{eq:wetterich} is exact, it is generally not possible to solve it and thus one has to resort to approximation schemes that render Eq.~\eqref{eq:wetterich} amenable to analytic and numerical calculations.
The first step in this direction is to use an Ansatz for the form of the effective action $\Gamma$ which, once inserted into Eq.~\eqref{eq:wetterich}, results in a set of coupled non-linear differential equations for the couplings which parametrize it. We consider the following LPA Ansatz~\cite{DUPUIS20211} for the modified effective action \(\Gamma_k\) of model A:
\begin{multline}
\label{eq:effective-action}
 \Gamma_k	 =  \Gamma_{0,k}[\tilde\phi_0,\phi_0] \\  + \int_{x} \vartheta(t-t_0)\res \left( Z_k \dot{\phi} + K_k \nabla^2\phi + \frac{\partial \mathcal{U}_k}{\partial \phi} - D_k \tilde\phi \right).
\end{multline}
The initial-time action  \(\Gamma_{0,k} = \Gamma_{0,k}[\tilde\phi_0,\phi_0]\) accounts for the initial conditions~\eqref{eq:initial-probability} and will be discussed later on. The field and time-independent factors \(Z_k, \ K_k\) and \(D_k\) in Eq.~\eqref{eq:effective-action} account for a possible renormalization of the derivatives and of the noise term, while the generic potential $\mathcal{U}_k=\mathcal{U}_k(\phi)$ encompasses the interactions of the model.

The potential \(\mathcal{U}_k=\mathcal{U}_k(\phi)\) is assumed to be  a \(\mathbb{Z}_2\)-symmetric local truncated polynomial, i.e.,
\begin{equation}\label{eq:AnsatzU}
\mathcal{U}_k(\phi)=\sum_{n=1}^{n_\text{tr}} \frac{g_{2n,k}}{(2n)!}(\phi^2-\bar{\phi}_k^2)^n,
\end{equation} 
where \(\bar{\phi}_k\) is a background field chosen to be the minimum of \(\mathcal{U}_k\) and \(n_\text{tr} \geq 2 \) is the truncation order. Every coupling $g_{2n,k}$ can be obtained from the expansion of the effective action as
\begin{equation}
\label{eq:vertexp-maintext}
g_{2n,k}  = \frac{\delta^{2n} \Gamma_k}{\delta\phi^{2n-1}\delta\res}\biggr|_{\substack{\res = 0 \\ \phi =\bar{\phi}}_k}, 
\end{equation}
where the derivatives of $\Gamma$ are evaluated at the homogeneous field configuration $\res =0 $ and $\phi = \bar{\phi}_k$. 
Finally, in order to derive the RG equations for the couplings appearing in the effective action~\eqref{eq:effective-action},  one has to take the derivative with respect to $k$ of both sides of Eq.~\eqref{eq:vertexp-maintext} and, by using Eq.~\eqref{eq:wetterich}, one finds 
\begin{multline}
\label{flowcoupling}
 \!\frac{\dd g_{2n,k}}{\dd k}  \!=\! \frac{\delta^{2n}}{\delta\phi^{2n-1}\delta\res} \frac{\dd \Gamma_k}{\dd k} \biggr|_{\substack{\res =0 \\ \phi =\bar{\phi}}_k} 
 \!\!\!\!+ \frac{\delta^{2n+1} \Gamma_k}{\delta\phi^{2n}\delta\res}\biggr|_{\substack{\res =0 \\ \phi =\bar{\phi}}_k} \frac{\dd \bar{\phi}_k}{\dd k},
\end{multline}
from which one can evaluate the flow equation for the couplings $g_{2n,k}$.

While in Ref.~\onlinecite{Chiocchetta2016} the exact form of \(\Gamma_{0,k}\) was required to determine the initial critical exponent \(\theta\) via a short-time analysis of the dynamics, here it will not play a major role since we are interested in the long-time limit of the aging regime.
In the following we introduce a different approach which does not make use of the explicit form of \(\Gamma_{0,k}\) in order to calculate \(\theta\): we implement the initial condition via the Dirichlet condition on the correlation function \eqref{Dirichlet}. This approach is justified since the fixed point of \(\tau_{0,k}\), because of its canonical dimension, is known to be \( \tau_{0}=+\infty\)~\cite{Janssen1989,Chiocchetta2016}.

In the following, we show how the LPA Ansatz simplifies the fRG equation~\eqref{eq:wetterich} for the modified effective action \(\Gamma_k\).
By taking advantage of the locality in space and time of the LPA Ansatz~\eqref{eq:effective-action}, i.e., of the fact that it is written as an integral over \(x\), one can rewrite the second of  Eqs.~\eqref{eq:Gamma-properties} as
\begin{equation}\label{preDyson}
G_k^{-1}(x,x')= G_{0,k}^{-1}(x,x') - \delta(x-x')\Sigma_k(x),
\end{equation} 
where we separated \(\Gamma_k^{(2)}+R_k\sigma\) in Eq.~\eqref{eq:Gamma-properties} in a field-independent and a field-dependent part \(G_{0,k}=G_{0,k}(x,x')\) and \(\Sigma_k=\Sigma_k(x)\), respectively. 
In order to derive \(G_{0,k}\) one should invert the field-independent part in Eq.~\eqref{eq:Gamma-properties}, while imposing Eq.~\eqref{Dirichlet} on the correlation function (the response function does not depend on the initial condition within the Gaussian approximation, see Sec.~\ref{sec:gaussian}). The analytical expression of \(G_{0,k}\) and \(\Sigma_k\) is reported in App.~\ref{subsec:app1-1}. 
Inverting Eq.~\eqref{preDyson} leads to a Dyson equation for \(G_k\):  
\begin{equation}
\label{Dyson}
G_k(x,x') = G_{0k}(x,x') + \!\! \int_{y}\! G_{0k}(x,y) \Sigma_k(y) G_k(y,x').
\end{equation}
One can then cast the fRG equation \eqref{eq:wetterich} in a simpler form by using  Eq.~\eqref{Dyson}. In fact, by formally solving the Dyson equation \eqref{Dyson}, one can express \(G_k\) as a series which, together with Eq.~\eqref{eq:wetterich}, renders 
\begin{equation}\label{eq:Wett2}
\frac{\dd\Gamma_k}{\dd k} = \sum_{n=1}^{+\infty} \Delta \Gamma_{n,k},
\end{equation}
where the functions \(\Delta\Gamma_{n,k}\) are given by 
\begin{multline}\label{eq:convolution}
\Delta \Gamma_{n,k} = \frac{1}{2}\int_{x,y_1 \cdots y_n} \text{tr} \Big[ G_{0,k}(x,y_1)\Sigma_k(y_1) 
 G_{0,k}(y_1,y_2)  \\ \left.\times \cdots \times \Sigma_k(y_n) G_{0,k}(y_n,x) \frac{\dd R_k}{\dd k} \sigma  \right].
\end{multline}
As a simple but instructive example, we consider a truncation of  the effective potential \(\mathcal{U}_k\) up to the fourth power of the field \(\phi\) (i.e., \(n_\text{tr}=2\)) around a  configuration with vanishing minimum \(\bar{\phi}_k=0\), i.e.,   
\begin{equation}
\label{Uphi0}
\mathcal{U}_k = \frac{r_k}{2}\phi^2 + \frac{g_k}{4!}\phi^4.
\end{equation}
The field-independent part of \(\Gamma^{(2)}\), i.e., \(-\Sigma_k\),  can be explicitly evaluated as
\begin{equation}
\label{sigmaf0}
\Sigma_k = - g_k \ \vartheta(t-t_0) \  \begin{pmatrix}
\displaystyle{\tilde \phi} \phi, &  \phi^2/2
\\ \phi^2/2, & 0 
\end{pmatrix}.
\end{equation}
Since \(\Sigma_k\) appears \(n\) times in the convolution \eqref{eq:convolution} which defines \(\Delta\Gamma_{n,k}\) on the r.h.s.~of Eq.~\eqref{eq:Wett2}, it follows that \(\Delta\Gamma_{n,k}\) contains a product of \(2n\) fields (grouped in \(n\)  pairs which depend on  different times \(t_i\) for \(i=1,...,n\)). 
Accordingly, the only terms on the r.h.s. of Eq.~\eqref{eq:Wett2} which contribute to renormalize $\mathcal{U}_k$ are \(\Delta\Gamma_{1,k}\) and \(\Delta\Gamma_{2,k}\).
Also the l.h.s. of Eq.~\eqref{eq:Wett2} is a polynomial of the fields, because of the LPA Ansatz \eqref{eq:effective-action}, and therefore each term of the expansion on the l.h.s. is uniquely matched by a term of the expansion on the r.h.s..
We note that \(\Delta\Gamma_{2,k}\), at variance with \(\Delta\Gamma_{1,k}\),  is a quantity that depends on the fields \(\phi\) and \(\tilde\phi\) evaluated at different times. Since the l.h.s. of Eq.~\eqref{eq:Wett2}, with the LPA Ansatz \eqref{eq:effective-action}, is local in time and space, one has to retain only local contributions in the r.h.s.
 In App.~\ref{subsec:app1-2} we detail how the  \(\Delta\Gamma_{i,k}\)'s terms with \(i=1,2\) can be calculated. In the following sections, we investigate how different truncations lead to time-dependent effective potential as an effect of the quench.

\subsection{Time-dependent effective action}
\label{sec:truncation-symmetric}
We derive here the RG equations from the Ansatz~\eqref{eq:effective-action} with the quartic potential $\mathcal{U}$ introduced in Eq.~\eqref{Uphi0}. 
Since this Ansatz corresponds to a local potential approximation~\cite{Berges2002, Delamotte2007}, the anomalous dimensions of the derivative terms ($K_k,Z_k$) and of the noise strength $D_k$ vanish, and therefore for simplicity,  we set $K_k = Z_k = D_k = 1$ in what follows.  Accordingly, the anomalous dimension $\eta$ and the dynamical critical exponent $z$ are equal to their Gaussian values $\eta = 0$ and $z =2$, respectively.

The only non-irrelevant terms which are renormalized within this scheme are those proportional to quadratic and quartic powers of the fields $\phi$ and $\res$, i.e., those associated with the post-quench parameter $r_k$ and the coupling $g_k$. 
The renormalization of the quadratic terms is determined by the contribution $\Delta \Gamma_{1,k}$ appearing on the r.h.s.~of Eq.~\eqref{eq:Wett2}, while the renormalization of the quartic one by the contribution $\Delta \Gamma_{2,k}$. 
The calculations of these two contributions are detailed in App.~\ref{sec:APP2}, and in particular App.~\ref{subsec:app1.3}, using an optimized cutoff function \(R_k\),  and leads to the following fRG flow equations:
\begin{subequations}
\label{eq:beta}
\begin{align}
\label{eq:r-eq}
\frac{\dd r_k(t)}{\dd k} &= -k^{d+1}\frac{a_d}{d}\frac{g_k }{\omega_k^2}[1-\ee^{-2\omega_kt}f_1(\omega_k t)],
\\
\label{eq:g-eq}
\frac{\dd g_k(t)}{\dd k } &= 6k^{d+1}\frac{a_d}{d}\frac{g_k^2}{\omega_k^3}[1-\ee^{-2\omega_kt}f_2(\omega_k t)],
\end{align}
\end{subequations}
where \(\omega_k=r_k +k^2\),  \(f_1(x)= 1+ 2x\), \(f_2(x)= 1+ 2x^2+2x\), $a_d = 2/[\Gamma(d/2) (4\pi)^{d/2}]$, \(d\) is the spatial dimensionality of the system and $\Gamma(x)$ is  the gamma function. Within the standard LPA approximation to non-equilibrium systems, the time dependence of \(r_k\) and \(g_k\) on the l.h.s. is exclusively given by the non-equilibrium initial condition~\eqref{eq:initial-probability} via \(\Delta\Gamma_{1,k}\) and \(\Delta\Gamma_{2,k}\) in Eq.~\eqref{eq:Wett2} and ultimately by \(G_{0,k}\), via the Dirichlet correlation function~\eqref{Dirichlet}. Remarkably, the time dependence of the couplings \(r_k\) and \(g_k\) vanishes exponentially in time:  for \(t\rightarrow\infty\)  one thus obtains the RG flow equations for the bulk part of the Ansatz~\eqref{eq:effective-action},  which determine the equilibrium fixed points. This structure is preserved for higher-order truncation in LPA approximation, as we detail in App.~\ref{subsec:app1-4}.

 In the long-time equilibrium regime, the fRG flow equations~\eqref{eq:beta} can be further simplified by rewriting them in terms of the dimensionless couplings \( \tilde r_k = r_k k^{-2}\) and \(\tilde g_k = (a_d/d) g_k  k^{4-d} \). The fixed points for the various values of the spatial dimensionality \(d\) are found by solving the system of equations corresponding to requiring vanishing derivatives of the dimensionless couplings, i.e.,  
\begin{subequations}
\label{eq:fixed point}
\begin{align}
\label{eq:r-eq-dimensionless}
& 0 = 2 \tilde r^* - \frac{\tilde g^*}{(1+\tilde r^*)^2}, \\ \vspace{2mm}
\label{eq:g-eq-dimensionless}
& 0 = (d-4)\tilde g^* + 6 \frac{\tilde g^{*2}}{(1+\tilde r^*)^3},
\end{align}
\end{subequations}
where the superscript \(^*\) indicates fixed-point quantities.
Equations~\eqref{eq:r-eq-dimensionless} and~\eqref{eq:g-eq-dimensionless} admit two solutions: the Gaussian fixed point $(\widetilde{r}_\text{G}^*,\widetilde{g}_\text{G}^*) = (0,0)$ and the Wilson-Fisher (WF) one, which, at leading order in $\epsilon = 4-d$, reads $(\widetilde{r}^*_\text{WF},\widetilde{g}^*_\text{WF})= (-\epsilon/12, \epsilon/6) + O(\epsilon^2)$.  By linearizing Eqs.~\eqref{eq:r-eq-dimensionless} and~\eqref{eq:g-eq-dimensionless} around these solutions, one finds that the Gaussian fixed point is stable only for $d >4$, while the WF fixed point is stable only for $d<4$. The latter has an unstable direction, and from the inverse of the negative eigenvalue of the associated stability matrix, one derives the critical exponent $\nu$, which reads $\nu=1/2+\epsilon/12 + O(\epsilon^2)$ and is the same as in equilibrium~\cite{Tauberbook2014}.

\subsection{The case with $\bar{\phi}_k  \neq 0$ }
\label{sec:truncation-ordered}
In this section, we consider the  approximation for the effective potential~\eqref{eq:AnsatzU} with a non-vanishing background.
This case differs from the one considered in Eq.~\eqref{Uphi0}, since it corresponds to an expansion around a finite homogeneous value $\bar{\phi}_k$: this choice allows one to capture the leading divergences of two-loop corrections in a calculation which is technically done at one-loop, as typical of  background field methods (see, e.g., Refs.~\onlinecite{Berges2002, Delamotte2007,Canet2007}); accordingly, one calculate, for instance, the renormalization of the factors $Z_k$, $K_k$, and $D_k$. In fact, the presence of the background field $\bar{\phi}_k$  reduces two-loop diagrams to one-loop ones in which an internal classical line (corresponding to a correlation function, $C_{0,k}$) has been replaced by the insertion of two expectation values $\bar{\phi}_k$. This is illustrated in the two diagrams below in which the external straight lines stand for the field $\phi$, while the wiggly lines indicate  the response field $\tilde{\phi}$; see also, e.g., Ref.~\onlinecite{Tauberbook2014}. For instance, in the case \(n_\text{tr}=2\), where Eq.~\eqref{eq:AnsatzU} can be conveniently rewritten as
\begin{equation}
\label{potentialordered}
\mathcal{U}_k(\phi) = \frac{g_k}{4!}\left(\phi^2-\bar{\phi}_k^2 \right),
\end{equation}
the renormalization of $Z_k$ and $K_k$ for  come from the diagram 
%
%
\begin{equation} \nonumber
\begin{tikzpicture}[baseline={([yshift=-.5ex]current bounding box.center)}]
\coordinate[label={[label distance=-1]above:$\mathcal{C}_{0}$}] (o) at (0,0);
\coordinate[label={[label distance=3.5]above left:$g$}] (ol) at (-0.5,0);
\coordinate[label={[label distance=3.5]above right:$g$}] (or) at (0.5,0);
\coordinate[] (oll) at (-1.25,0);
\coordinate[] (orr) at (1.25,0);
\coordinate[label={[label distance=-1]above:$\mathcal{C}_{0}$}] (on)  at (0,0.5);
\coordinate[label={[label distance=0]below:$\mathcal{R}_{0}$}] (on)  at (0,-0.5);
\draw[thick] (or) -- (ol);
\draw[res] (or) arc (0:-90:0.5);
\draw[thick] (or) arc (0:270:0.5);
\draw[res] (ol) node[circle,fill,inner sep=1pt]{} -- (oll);
\draw[thick] (or) node[circle,fill,inner sep=1pt]{} -- (orr);
\end{tikzpicture}
\quad  \xrightarrow[]{\text{fRG}}
\begin{tikzpicture}[baseline={([yshift=-.5ex]current bounding box.center)}]
\coordinate[] (o) at (0,0);
\coordinate[label={[label distance=3.5]left:$g_k \ $}] (ol) at (-0.5,0);
\coordinate[label={[label distance=3.5]right:$ \ g_k$}] (or) at (0.5,0);
\coordinate[label={[label distance=1]above left:$\bar{\phi}_k$}] (no) at (-1.25,0.5);
\coordinate[label={[label distance=1]above right:$\bar{\phi}_k$}] (ne) at (1.25, 0.5);
\coordinate[] (so) at (-1.25,-0.5);
\coordinate[] (se) at (1.25,-0.5);
\coordinate[label={[label distance=-1]above:$\mathcal{C}_{0,k}$}] (on)  at (0,0.5);
\coordinate[label={[label distance=0]below:$\mathcal{R}_{0,k}$}] (on)  at (0,-0.5);
\draw[res] (or) arc (0:-90:0.5);
\draw[thick] (or) arc (0:270:0.5);
\draw[thick] (ol) node[circle,fill,inner sep=1pt]{} -- (no) node[draw,cross out,rotate=60]{};
\draw[res] (so) -- (ol);
\draw[thick] (ne) node[draw,cross out,rotate=30]{} -- (or);
\draw[thick] (or) node[circle,fill,inner sep=1pt]{} -- (se);
\end{tikzpicture}, 
\end{equation}
while the renormalization of the noise strength $D_k$ comes from the diagram:
\begin{equation} \nonumber
\begin{tikzpicture}[baseline={([yshift=-.5ex]current bounding box.center)}]
\coordinate[label={[label distance=-1]above:$\mathcal{C}_{0}$}] (o) at (0,0);
\coordinate[label={[label distance=3.5]above left:$g$}] (ol) at (-0.5,0);
\coordinate[label={[label distance=3.5]above right:$ g$}] (or) at (0.5,0);
\coordinate[] (oll) at (-1.25,0);
\coordinate[] (orr) at (1.25,0);
\coordinate[label={[label distance=-1]above:$\mathcal{C}_{0}$}] (on)  at (0,0.5);
\coordinate[label={[label distance=0]below:$\mathcal{C}_{0}$}] (on)  at (0,-0.5);
\draw[thick] (or) -- (ol);
\draw[thick] (or) arc (0:-90:0.5);
\draw[thick] (or) arc (0:270:0.5);
\draw[res] (ol) node[circle,fill,inner sep=1pt]{} -- (oll);
\draw[res] (or) node[circle,fill,inner sep=1pt]{} -- (orr);
\end{tikzpicture}
\quad  \xrightarrow[]{\text{fRG}}
\begin{tikzpicture}[baseline={([yshift=-.5ex]current bounding box.center)}]
\coordinate[] (o) at (0,0);
\coordinate[label={[label distance=3.5]left:$g_k \ $}] (ol) at (-0.5,0);
\coordinate[label={[label distance=3.5]right:$ \ \ g_k$}] (or) at (0.5,0);
\coordinate[label={[label distance=1]above left:$\bar{\phi}_k$}] (no) at (-1.25,0.5);
\coordinate[label={[label distance=1]above right:$\bar{\phi}_k$}] (ne) at (1.25, 0.5);
\coordinate[] (so) at (-1.25,-0.5);
\coordinate[] (se) at (1.25,-0.5);
\coordinate[label={[label distance=-1]above:$\mathcal{C}_{0,k}$}] (on)  at (0,0.5);
\coordinate[label={[label distance=0]below:$\mathcal{C}_{0,k}$}] (on)  at (0,-0.5);
\draw[thick] (or) arc (0:-90:0.5);
\draw[thick] (or) arc (0:270:0.5);
\draw[thick] (ol) node[circle,fill,inner sep=1pt]{} -- (no) node[draw,cross out,rotate=60]{};
\draw[res] (so) -- (ol);
\draw[thick] (ne) node[draw,cross out,rotate=30]{} -- (or);
\draw[res] (or) node[circle,fill,inner sep=1pt]{} -- (se);
\end{tikzpicture}.
\end{equation}
In these diagrams, those on the right indicate how the corresponding perturbative diagrams encountered in pRG, reported on the left, are reproduced within fRG (see, e.g.,  Ref.~\onlinecite{Chiocchetta2016})

The flow equations for $Z_k$ and $K_k$ can be conveniently expressed in terms of the corresponding anomalous dimensions $\eta_D$, $\eta_Z$ and $\eta_K$, defined as 
\(\eta_D \equiv -\dd \log D_k /\dd \log k \) and similarly for the others.
We note that $\eta_D = \eta_Z$: this a consequence of detailed balance~\cite{Tauberbook2014,Canet2007}, which characterizes the equilibrium dynamics of model A. In fact, while the short-time dynamics after the quench violates detailed balance inasmuch as time-translational invariance is broken, in the long-time limit (in which the flow equations are valid) detailed balance is restored.
From general scaling arguments~\cite{Tauberbook2014}, the static anomalous dimension \(\eta\) and the dynamical critical exponent \(z\) are given by
\begin{equation}
\label{anomalousdimension}
\eta = \eta^*_K,  \quad  \quad \quad
 z = 2- \eta^*_K + \eta_Z^*.
\end{equation}
The calculation of the anomalous dimension $\eta$ and the dynamic critical exponent \(z\), discussed in Ref.~\onlinecite{Chiocchetta2016} (to which we refer the reader for details), gives these anomalous dimensions in terms of the fixed-point values of the couplings which define the LPA Ansatz.

Let us discuss how the RG equations can be derived in the presence of a non-vanishing background field $\bar{\phi}_k$.
First, Eq.~\eqref{flowcoupling} shows that the \(k\)-derivative of \(\bar{\phi}_k\) enters the fRG flow equations for the couplings~\eqref{eq:vertexp-maintext}. In order to define unambiguously \(\bar{\phi}_k\), it is possible to change the variable \(\phi\), which the effective potential \(\mathcal{U}_k\) depends on, to the \(\mathbb{Z}_2\)-invariant \(\rho = \phi^2/2\). One then defines the minimum as the configuration which corresponds to a vanishing \(\rho\)-derivative of \(\mathcal{U}_k(\rho) = \mathcal{U}_k(\phi^2/2)\). The computations that follows in order to obtain the \(k\)-derivative of \(\bar{\phi}_k\) are detailed in App.~\ref{subsec:app1-5}. There, it is shown how the fRG flow equation for \(\bar{\phi}_k\) is given in terms of \(\Delta\Gamma_{1,k}\), while the one for \(g_k\) is given in terms of \(\Delta\Gamma_{2,k}\). Finally, using the computation of these terms, detailed in App.~\ref{subsec:app1-2}, one determines the fRG equation of the parameters which define the Ansatz \eqref{potentialordered}.   
The time dependence of these fRG flow equations enters via decreasing exponentials, as in  Eqs.~\eqref{eq:beta}.  Accordingly, this allows a straightforward analysis of the equilibrium fixed point of the Ising universality class, as done for the case of vanishing background field approximation. This consideration still holds true for the case of higher-order truncation, as explained in App.~\ref{subsec:app1-4}.

The fixed point can be determined numerically, similarly to what is done in Eq.~\eqref{eq:fixed point}, once the fRG equation for the coupling \(g_k\) and the background field \(\phi_k\) are accompanied by that for the derivative parameter \( K_k\), to which they are coupled \cite{Chiocchetta2016}  and by taking advantage of the relation \(\eta_D = \eta_Z\).

\section{Aging regime}
\label{sec:5}
In this section we discuss how to obtain the two-point correlation and response functions of the order parameter within the LPA approximation, detailing how the aging regime~\eqref{eq:greens-scaling} is recovered.

\subsection{Two-time functions in LPA}
Let us denote by $\mathcal{G}(t,t')$ the \(2\times 2\) matrix of the physical two-point correlations (cf. App.~\ref{sec:APP2}),
whose entries are given by the response \(\mathcal{R}\) and correlation \(\mathcal{C}\) functions, defined in Eqs.~\eqref{defR} and \eqref{defC}, respectively. $\mathcal{G}(t,t')$ can be calculated as the inverse of the physical effective action $\Gamma_{k=0}\equiv \Gamma$, evaluated in the configuration of \(\Phi\) minimizing \(\Gamma\) (see Eq.~\eqref{eq:Gamma-properties} and Ref.~\onlinecite{Tauberbook2014}).
This minimum configuration is given by \(\Phi = 0\) for the critical quench, since the background field $\bar{\phi}_k$ vanishes at $k=0$, as a consequence of criticality. Moreover, within the LPA, \(\Gamma^{(2)}\) is assumed to be a local quantity in space and time, i.e., \(\Gamma^{(2)}(x,x')=\delta(x-x')\Gamma^{(2)}(x)\). 
With these two simplifications, the equation of motion for the physical two-point function \(\mathcal{G}(t,t')\)  becomes
\begin{equation}\label{eq:motionlocal}
\left.\Gamma^{(2)}_\bold{q}(t)\right|_{\Phi= \bold{0}}\mathcal{G}_\bold{q}(t,t') =\delta(t-t'),
\end{equation} 
where we exploited the space-translational invariance of model A, and Fourier-transformed with respect to this spatial dependence.
 
In order to retrieve the two-point function \(\mathcal{G}\) from Eq.~\eqref{eq:motionlocal}, one should first compute \(\Gamma^{(2)}|_{\Phi=0}\) given the LPA Ansatz \eqref{eq:effective-action} for the effective action.  By implementing the Dirichlet boundary condition~\eqref{Dirichlet}, the equation of motion~\eqref{eq:motionlocal} can be cast in the form of a set of equations for the response and correlation functions. Thus, one obtains
\begin{subequations}
\label{eq:greens-motion-lpa}
\begin{align}
& \label{Rreduced}[Z \partial_t + r(t)]  \mathcal{R}_\bold{q=0}(t,t') = \delta(t-t'), \\ \vspace{5mm}
&  \label{Creduced} \mathcal{C}_\bold{q=0}(t,t') = 2 D \int^{\infty}_{t_0} ds \ \mathcal{R}_\bold{q=0}(t,s) \mathcal{R}_\bold{q=0}(t',s),
\end{align}
\end{subequations}
where $r(t)$ is defined as the $k\to 0$ limit of the renormalized parameter $r_k$ defined by
\begin{equation}
\label{defmass}
  r_k \equiv \frac{\partial^2 \mathcal{U}_k}{\partial \phi^2} (\phi= 0) \simeq 
  \begin{cases}
\  r_\Lambda & \text{for} \quad k\rightarrow \Lambda, \\
\  r(t) & \text{for} \quad k\rightarrow 0.
\end{cases}
\end{equation}  
The time dependence of \(r(t) = r_{k=0}(t)\) in Eq.~\eqref{eq:greens-motion-lpa} is given by the fRG flow equation, as shown, for instance, in Eq.~\eqref{eq:r-eq}. 
In order to access the aging regime, one requires the critical point to be reached at long times, i.e., $r(t) \to 0$ for $t \to +\infty$. This condition can be achieved by properly fine-tuning the microscopic value \(r_\Lambda\), as discussed  in more detail in Sec.~\ref{subsec:Green}.

\subsection{Long-time aging dynamics}
\label{subsec:Green}

We first consider the case \(\bar \phi_k =0\) and \(n_\text{tr}=2\) discussed in Sec.~\ref{sec:truncation-symmetric}. We consider the case of a critical quench and we denote by $r^*(t)$ the fine-tuned value of $r(t)$ which vanishes in the long-time limit.
By integrating Eq.~\eqref{eq:r-eq} in $k$ from the microscopic scale \(\Lambda\) to \(k=0\) and evaluating the couplings \(r_k\) and \(g_k\) at their fixed point values, one obtains
\begin{equation}\label{r}
r^*(t) =  - \frac{\theta}{t}\left[1-e^{-2\Lambda^z t (1+\tilde r^*)} \tilde{F}_1(\Lambda^z t) \right],
\end{equation}
with \(\tilde{F}_1(x) = 1+x(1+\tilde r^*)\).  The value of \(\theta\), the meaning of which  will be discussed further below, is given by
\begin{equation}\label{thetalpa}
\theta = \frac{\tilde g^*}{z(1+\tilde r^*)^2},
\end{equation}
with \(z=2\) and the fixed-point values \(\tilde r^*\) and \(\tilde g^*\) given by the solution of Eqs.~\eqref{eq:fixed point}.
In order to obtain Eq.~\eqref{r}, the microscopic value \(r_\Lambda\) has been fine-tuned to \(r_\Lambda^*=-\Lambda^z \tilde g^*(1+\tilde r^*)^{-2}/z\), implying that the critical temperature is shifted towards a smaller value compared to the Gaussian one, \(r^*_\Lambda=0\), as expected because of the presence of fluctuations~\cite{Tauberbook2014}.

\begin{figure}[!t]

\hspace{2cm}
\includegraphics[scale=0.6, trim = 8cm 10cm 1cm 0cm]{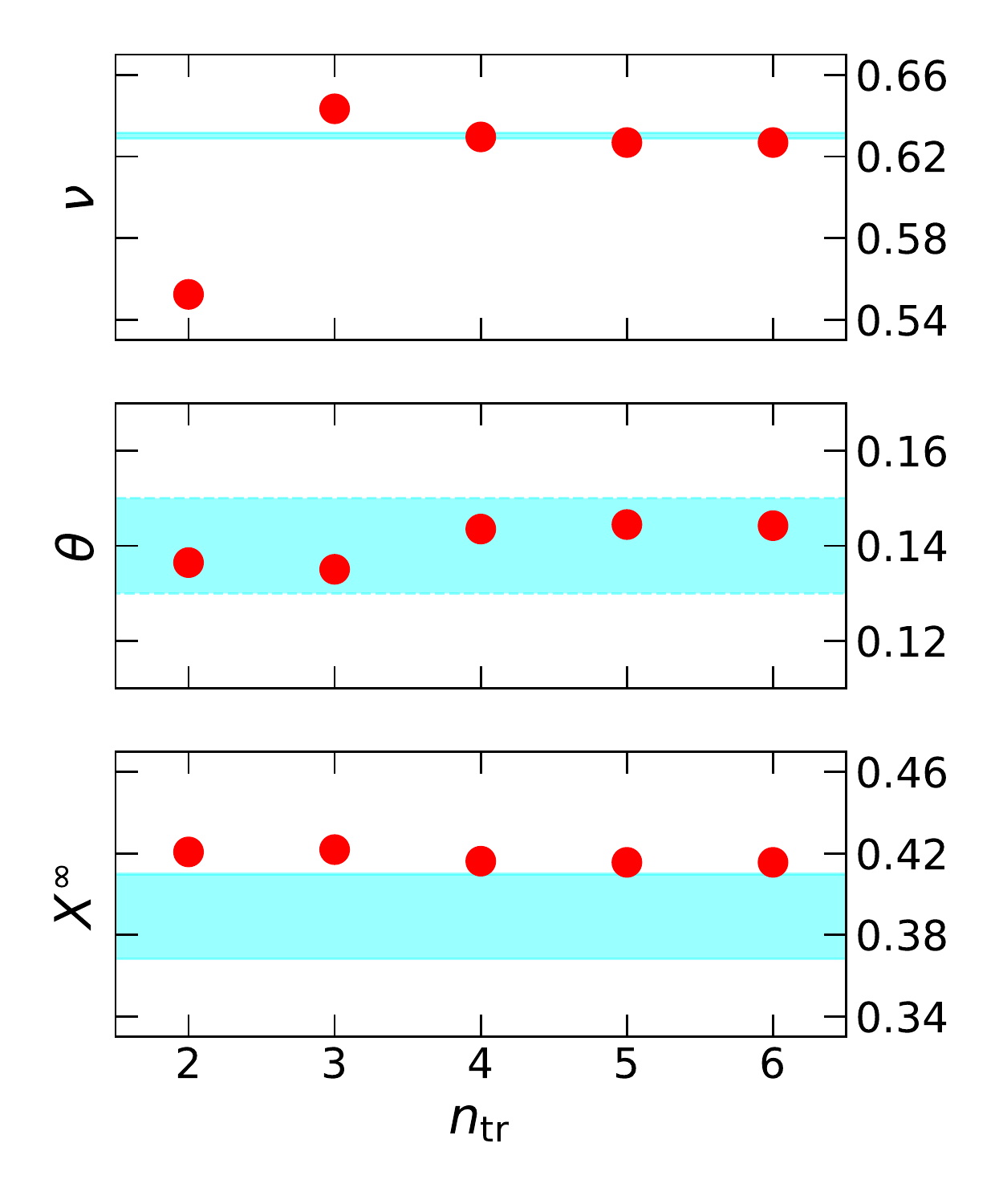}
\vspace{5.3cm}
\caption{ Numerical estimates of the equilibrium critical exponent \(\nu\), of the non-equilibrium critical initial-slip exponent \(\theta\) and of the universal amplitude ratio \(X^\infty\) in spatial dimension \(d=3\), as functions of the truncation order \(n_\text{tr}\).  Red dots, instead, correspond to the non-vanishing background field approximation for \(\mathcal{U}_k\) discussed in Sec.~\ref{sec:truncation-ordered}. The shaded cyan areas indicate the numerical estimates derived from the Monte Carlo simulations available in the literature.}
\label{fig:trunc}
\end{figure}

Let us now solve the equations of motion for the two-time functions given in Eqs.~\eqref{eq:greens-motion-lpa}. Since we are solely interested in the aging regime, we shall assume \(t \gg t' \gg \Lambda^{-2}\). In this case, one can use the asymptotic value of the parameter $r^*(t) $ given by
\begin{equation}
\label{raging}
r^*(t) = -\theta/t,
\end{equation} 
as detailed in App.~\ref{sec:proof}, and the solutions reads:
\begin{equation}
\label{scalingLPA}
\!\mathcal{R}_\bold{q=0}(t,t')  =\left(\frac{t}{t'}\right)^\theta\!\!\!,  \text{ and}\,\,\, 
\mathcal{C}_\bold{q=0}(t,t')  =  \frac{2 t'}{1-2\theta}   \left(\frac{t}{t'}\right)^{\theta}\!\!\!. 
\end{equation}
By comparing Eqs.~\eqref{scalingLPA}  with their scaling form \eqref{eq:greens-scaling}, one finds that the non-universal amplitudes of the response and the correlation functions are given by \(A_\mathcal{R}=1\), \(A_\mathcal{C}= 2 (1-\theta)^{-1}\), respectively, and that \(\theta\) introduced in Eq.~\eqref{thetalpa} is nothing but the critical initial-slip exponent. This method to derive $\theta$ is alternative to that employed in Ref.~\onlinecite{Chiocchetta2016}, which was based on the renormalization of the initial-time action $\Gamma_0$.
Our result for \(\theta\) exactly matches the one obtained in Ref.~\onlinecite{Chiocchetta2016} within the same Ansatz for \(\mathcal{U}_k\), as explained in App.~\ref{comparison}. 

The expression of the FDR \(X^\infty\), given in Eq.~\eqref{Xampliratio} in terms of $A_R, \ A_C$, and \(\theta\), finally reads:
\begin{equation}\label{Xlpa}
X^\infty = \frac{1/2-\theta}{1-\theta}.
\end{equation} 
Accordingly, \(X^\infty\) is given only in terms of \(\theta\), which, in turn is  fully determined by the fixed points of Eqs.~\eqref{eq:fixed point}. We note that, from Eq.~\eqref{thetalpa} and \eqref{Xlpa}, one obtains \(\theta = \epsilon/6 +\mathcal{O}(\epsilon^2)\) and \(X^\infty= 1-\epsilon/12+\mathcal{O}(\epsilon^2)\) to leading order in  \(\epsilon=4-d\), thus retrieving the known one-loop result~\cite{Calabrese2005}.
The validity of Eq.~\eqref{Xlpa} goes beyond the simple Ansatz  \(\bar{\phi}_k=0\) discussed in this section. In fact, it is a general consequence of the LPA approximation and it does not depend on the specific truncation of the potential \(\mathcal{U}_k\), as we prove in the next section. Accordingly, different LPA truncations will affect the specific value assumed by \(X^\infty\) only through the value of \(\theta\), as we discuss in what follows for a different truncation.

\subsection{Finite background field \(\bar{\phi}_k\neq 0 \)}
\label{ultima}
 
In the following, we discuss the case of the LPA Ansatz with the generic effective potential \(\mathcal{U}_k\) given by Eq.~\eqref{potentialordered}. The parameter $r_k$ defined in Eq.~\eqref{defmass}, is now given by
\begin{equation}\label{high-temp-mass}
r_k  = -\frac{1}{3!} g_k \bar{\phi}^2_{k}.
\end{equation}
The form of \(\dd r_k /\dd k\) is now determined by the flow equation of the parameters \(g_k\) and \(\bar{\phi}^2_{k}\), dicussed in Sec.~\ref{sec:truncation-ordered}.
The equation for the two-point function~\eqref{eq:greens-motion-lpa} can be cast in the following useful form
\begin{subequations}\label{eq:motionlpa'}
\begin{align}
&\left[\partial_t + \frac{r(t)}{Z}\right]\widetilde{\mathcal{R}}_\bold{q=0}(t,t')  =\delta(t-t'), \\
& \widetilde{\mathcal{C}}_\bold{q=0}(t,t')  = 2 \frac{D}{Z} \int_0^\infty ds  \ \widetilde{\mathcal{R}}_\bold{q=0}(t,s) \widetilde{\mathcal{R}}_\bold{q=0}(t',s),
\end{align}
\end{subequations}
with \(Z_k\) assumed to be time-independent, \(\widetilde{\mathcal{R}}_{\bold{q}}(t,t') = Z \mathcal{R}_{\bold{q}}(t,t')\), and \(\widetilde{\mathcal{C}}_{\bold{q}}(t,t')=Z\mathcal{C}_{\bold{q}}(t,t')\). 
The same derivation as in Sec.~\ref{subsec:Green} applies, upon replacing $r_k$  with \(r_k /Z_k\) in Eqs.~\eqref{eq:motionlpa'}. A canonical power counting shows that \(r/Z\) has the dimension of an inverse time, and therefore equations similar to Eqs.~\eqref{r} and \eqref{raging} can be obtained, as detailed in App.~\ref{sec:Apptheta}.

\begin{table}[!t]
\begin{ruledtabular}
\begin{tabular}{lll}
\text{Method} 	& \textbf{ \(\theta\)} & \textbf{\(X^\infty\)}  \\  \hline
fRG [this work]   				& 0.144		&  0.415   \\  
MC  & 0.14(1)~\cite{Grassberger1995}    	&   0.40(1)~\cite{Godreche2000}  \\    
MC  & 0.135(1)~\cite{Jaster1999}  ~  	&   0.380(13)~\cite{Prudnikov2017}  \\ 
MC  	&  				&   0.391(12)~\cite{Prudnikov2017}  \\    
pRG~  & 0.135(1) ~\cite{Prudnikov2008} 	&  0.429(6)~\cite{Calabrese2002}  \\   
\end{tabular}
 \caption{Summary of available estimates for the critical initial-slip exponent and the asymptotic value  \(X^\infty\) of the FDR, for \(d=3\).}
\label{tab:2}
\end{ruledtabular}
\end{table}
\begin{table}[!t]
\begin{ruledtabular}
\begin{tabular}{lll}
					& pRG at order \(\epsilon\) & fRG in LPA\\  \hline
$ A_R$			& $1+\mathcal{O}(\epsilon^2)$ & 1 \\
$ A_C$ 		& $2(1+2\theta)+\mathcal{O}(\epsilon^2)$ & $2/(1-2\theta)$  \\
$ X^\infty$	& $(1-\theta)/2+\mathcal{O}(\epsilon^2)$ & $(1/2-\theta)/(1-\theta)$\\   
\end{tabular}

      \caption{Comparison of predictions of the pRG treatment with those of fRG within the LPA approximation, for the various relevant quantities.}
      \label{tab:table1}
\end{ruledtabular}
\end{table}

The final prediction for \(\theta\) is different from Eq.~\eqref{thetalpa}, but it agrees order by order with the one found previously by means of the short-time analysis of the dynamics in Ref.~\onlinecite{Chiocchetta2016}, as we detail in App.~\ref{comparison}. 
By solving Eqs.~\eqref{eq:motionlpa'} for the reduced response and correlation function, $\tilde{R}$ and $\tilde{C}$, respectively, one obtains a form similar to the one given in Eq.~\eqref{scalingLPA} in the aging regime, using the fact that \(D_k/Z_k = \textrm{const.}\), which is a consequence of the fact that the flow equation of $D_k$ and $Z_k$ are identical~\cite{Chiocchetta2016} (the constant can then be reabsorbed in a suitable renormalization of the fields).
We find consequently that Eq.~\eqref{Xlpa} encompasses also the case in which the potential \(\mathcal{U}_k\) is expanded around its non-vanishing minimum. 

We emphasize that, because of the definition~\eqref{defmass}, the value of $r_k$ depends on the order $n_{\text{tr}}$ of the truncation in the Ansatz~\eqref{eq:AnsatzU}. For instance, for \(n_{\text{tr}}=3\), it is given by
\begin{equation}\label{massn3}
r_k =  -\frac{1}{3!} g_k \bar{\phi}^2_{k}+\frac{1}{5!} \lambda_{k} \bar{\phi}^4_{k},
\end{equation}
where \(\lambda_k = g_{6,k}\). In Ref.~\onlinecite{Chiocchetta2016} the term proportional to \(\lambda_k\) in Eq.~\eqref{massn3} was neglected, resulting in an inconsequential discrepancy in the value of \(\theta\) of \(2\%\) for \(d= 3\) compared to the one obtained here (see App.~\ref{formula}  for a detailed  comparison between the results of Ref.~\onlinecite{Chiocchetta2016} and those presented here).

\begin{figure*}[!t]
\hspace*{2cm}
\includegraphics[scale=0.32,trim = 12.5cm -2cm 3cm -1cm]{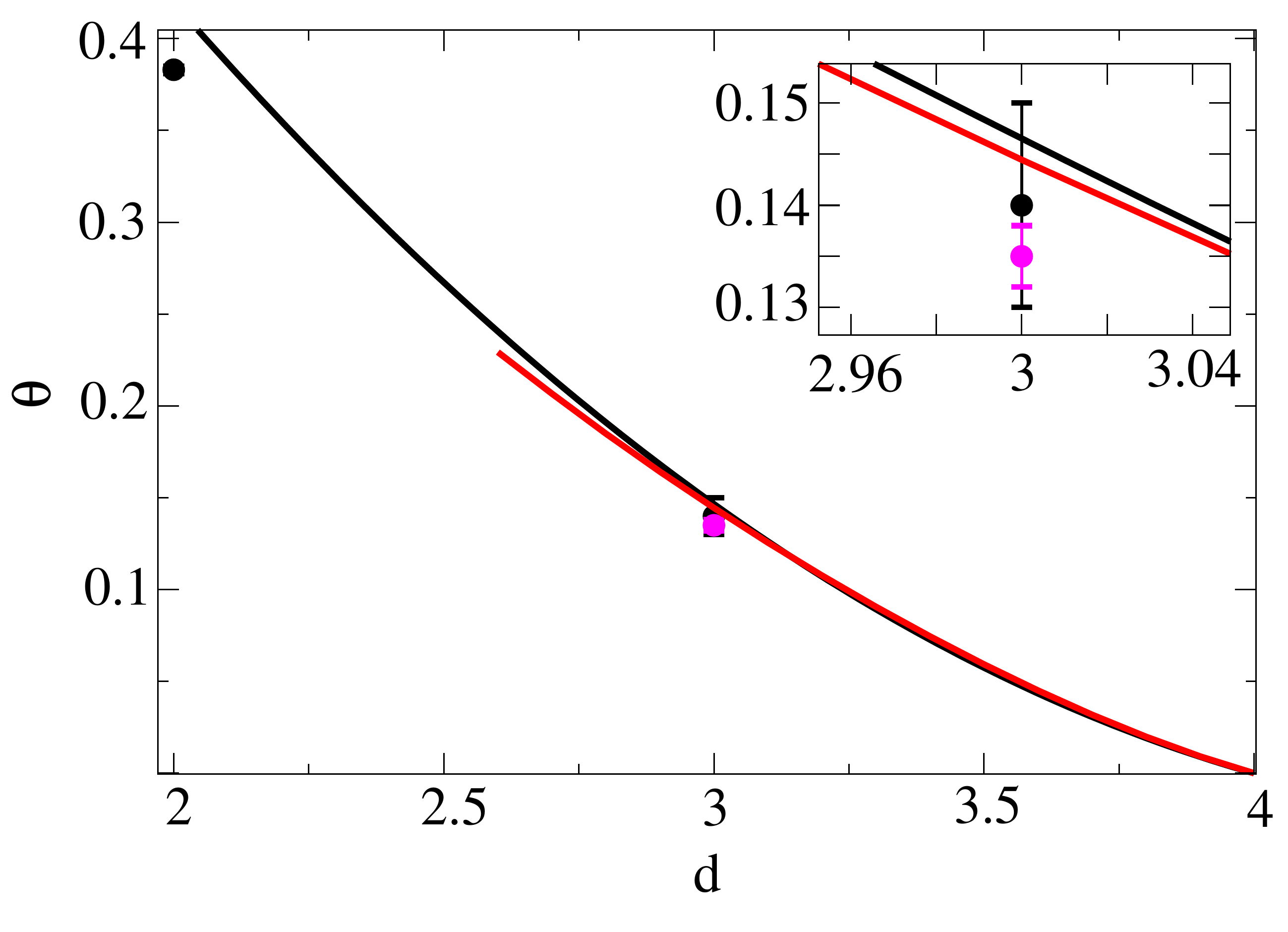}
\includegraphics[scale=0.32,trim = -0.5cm -2cm 5cm -1cm]{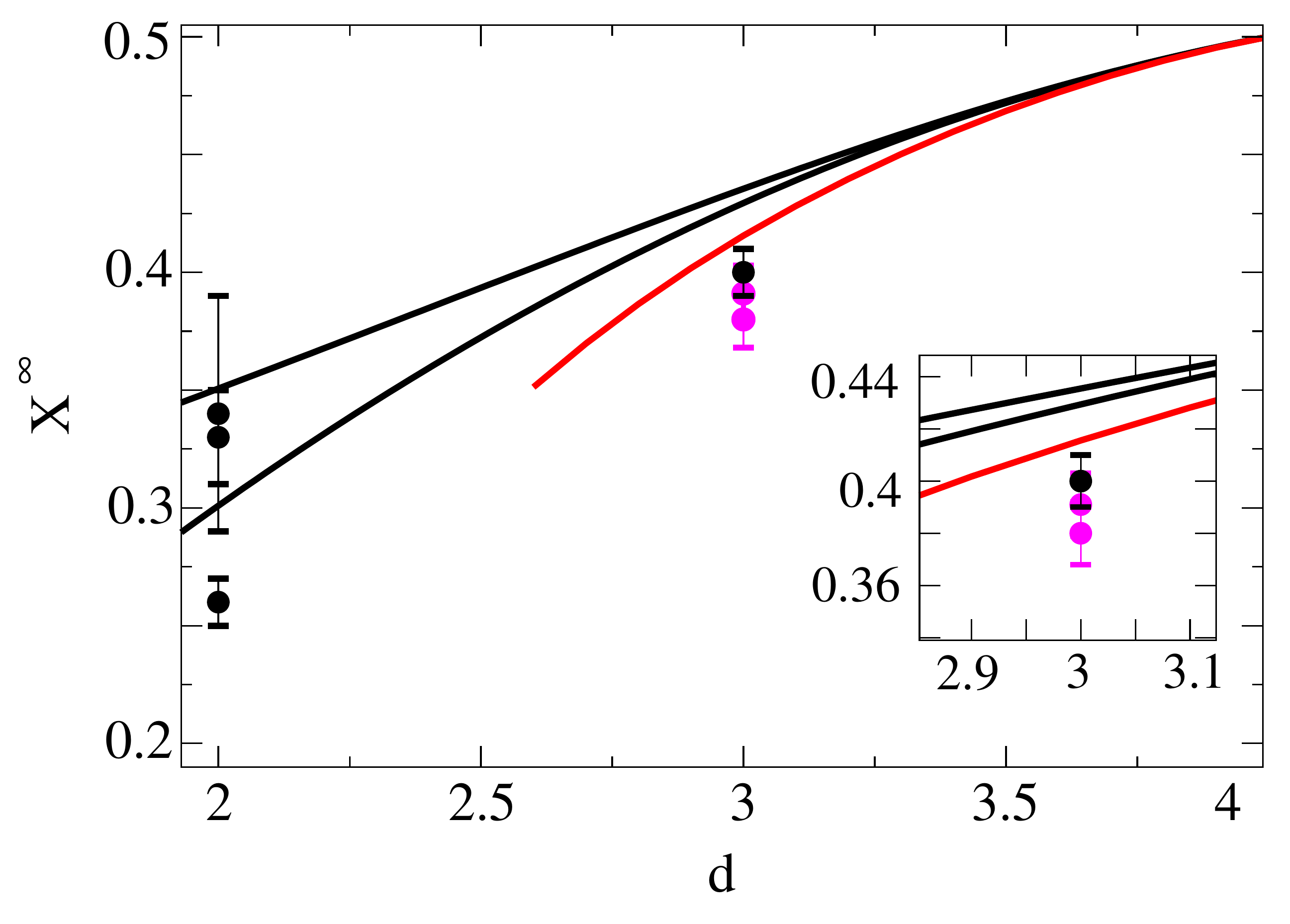}%
\hspace*{5cm}
\hspace*{0cm}
\vspace{-0.5cm}
\caption[]{\small Estimates of the critical initial-slip exponent \(\theta\) (left panel) and of the asymptotic value \(X^\infty\) of the FDR  (right panel) as a function of the spatial dimensionality \(d\) of the system. Red lines are obtained with the non-vanishing background field LPA approximation of the fRG, with \(n_{\text{tr}}=6\).  The black lines indicate Borel resummations of, respectively, three-loop (left panel, from Ref.~\onlinecite{Prudnikov2008}) and two-loop (right panel, from Ref.~\onlinecite{Calabrese2005}) pRG. 
For $d=3$ the Monte Carlo results for $\theta$ are from Refs.~\onlinecite{Grassberger1995} (black) and~\onlinecite{Jaster1999} (magenta), while for $X^\infty$ are from Refs.~\onlinecite{Godreche2000} (black) and~\onlinecite{Prudnikov2017} (magenta).
For $d=2$, all the Monte Carlo data are derived from Ref.~\onlinecite{Calabrese2005}. 
Insets: magnification of the main plots for \(d \simeq 3\).} \label{fig}
\end{figure*} 

\section{Discussion of the results}
\label{sec:results}

Here we discuss our predictions for the values of the critical initial-slip exponent $ \theta$ and for the universal amplitude ratio $X^\infty$.

We first consider the convergence of the predictions in Eqs.~\eqref{thetalpa} and~\eqref{Xlpa} upon increasing the order $n_\text{tr}$ of the truncation in Eq.~\eqref{eq:AnsatzU}. In Fig.~\ref{fig:trunc} we report the estimates of $\nu$,  $\theta$, and $X^\infty$  for $d=3$, obtained with a non-vanishing background field, for various  truncation orders $n_\text{tr}$. The values (red dots) display a rather small variation upon increasing $n_\text{tr}$, suggesting that their values have effectively converged to those corresponding to a full LPA solution. This fast convergence is due to the use of the optimized cutoff given by  Eq.~\eqref{Litim}, as pointed out in Ref.~\onlinecite{Litim2002}. While the values of $\nu$ and $\theta$ appear to be in very good agreement with the available MC data (shaded areas), the values of $X^\infty$ lie outside (though close to) the MC estimates. 
In Tab.~\ref{tab:2} we report the values of our best approximation \(n_\text{tr}=6\) for  \(d=3\) of $\theta$ and $X^\infty$, comparing them with the MC and pRG estimates available in literature.  For what concerns the values given by pRG, we report for \(\theta\) the result of the three-loop Borel summation procedure~\cite{Prudnikov2017}, while for \(X^\infty\) we report the low Pad\'{e} approximation of the two-loop calculation~\cite{Calabrese2002}. 

In Fig.~\ref{fig}, we report the predictions for $\theta$ and $X^\infty$ obtained with \(n_\text{tr}=6\) (red solid curve) as  functions of the spatial dimensionality $d$, and we compare them with MC (symbols) estimates and pRG  predictions (black curves). The agreement between fRG and pRG predictions for $\theta$ is remarkable for \(d \gtrsim 3.2\), while increasing discrepancies emerge at smaller values of \(d\).
The fRG predictions cannot be extended down to $d=2$ with the Ansatz considered. In fact, for \(d\leq3\) additional stable fixed points appear beyond the Wilson-Fisher one, while for \(d \leq 2.5\) the latter disappears. This is not surprising, since for \(d<3\) there exist more relevant terms than those considered in the truncation, which therefore is no longer sufficient and justified.

Concerning the values of  \(X^\infty\), the predictions of fRG correspond to that of pRG  close to $d=4$,  but it departs quite soon from it. The fRG prediction is much closer to the MC estimate in $d=3$ than the pRG one, although still not within its error bars.  

The significant improvement provided by the fRG compared to the pRG can be traced back to two reasons. First, the LPA approximation provides a resummation of one-loop diagrams, resulting in a more precise determination of the amplitudes $A_R$ and $A_C$, compared to the pRG predictions at order \(\epsilon\);
 a comparison of the analytical expressions between pRG and fRG at order \(\epsilon\) in LPA approximation is presented in Tab.~\ref{tab:table1}. Second, the non-perturbative nature of the fRG provides a more accurate estimate of $\theta$ (see Fig.~\ref{fig}, left panel) and therefore of \(X^\infty\).

\section{Conclusions and perspectives}
\label{sec: conclusions}
In this work we presented an approach, based on the functional renormalization group (fRG), to predict the universal fluctuation-dissipation ratio (FDR) \(X^\infty\) for the aging dynamics of model A. By calculating the two-time correlation functions from the renormalized effective action, we showed that, within the local potential approximation (LPA), the universal ratio \(X^\infty\) depends only on the critical initial-slip exponent \(\theta\). As a side result, we proposed an alternative way to calculate $\theta$ within the fRG by focusing on the long-time behavior of the two-time response and correlation functions. Moreover, we showed that the values of $\theta$ and  $X^\infty$ converge quickly upon increasing the order of the truncation in the Ansatz for the potential. Finally, we compared our prediction with the existing estimates obtained from Monte Carlo (MC) simulations and from the perturbative renormalization group (pRG), finding that the fRG results are closer to the MC than those of pRG ones.

Since the value of $X^\infty$ in $d=3$ is still not compatible with the MC results (although closer to it than predicted by previous approaches), a natural future direction is the computation of the two-time correlation functions beyond the LPA. Non-local terms are in fact expected to provide sizeable corrections, as they contribute with a term proportional to \(\epsilon^2\) in the pRG~\cite{Calabrese2002}.
This could be achieved by using different approximations for closing the infinite hierarchy of equations generated by the fRG equation, following, for instance, the approach of Ref.~\onlinecite{Kloss2012} or considering mode-coupling approximations~\cite{Bouchaud1996}.

Other natural directions consist in applying our method to characterize the FDR of other static universality classes, e.g., $O(N)$ and Potts models, or dynamics with conserved quantities~\cite{Oerding1993, Oerding1993b}. Moreover, the extension of the FDR analysis to non-equilibrium quantum systems~\cite{Langen2016,Sieberer2013,Altman2015,Nicklas2015,Chiocchetta2017,Pruefer2018,Erne2018} represents an intriguing issue.

\begin{acknowledgments}
We are indebted to R. Ben Alì Zinati and M. Scherer for useful discussions. A.~C. acknowledges support by the funding from the European Research Council (ERC) under the Horizon 2020 research and innovation program, Grant Agreement No. 647434 (DOQS). 
{M.~V. warmly thanks the University of Milan and SISSA for their hospitality during the first phases of this project.}
\end{acknowledgments}

\appendix

\section{Effective action}
\label{effective action}

Starting from the \(k\)-dependent action \(S_k\), we define the associated generating function $W_k[J]$ as
\begin{equation}
\label{eq:W-definition}
W_k[J] = \log \left[\int \mathrm{D}\Psi\, \ee^{-S_k[\Psi] + \int_{t,\rr}\Psi^\text{t} J}\right],
\end{equation}
where $\mathrm{D}\Psi$ denotes functional integration over both the fields $\varphi$ and $\widetilde{\varphi}$, while $J^\text{t}=(j,\widetilde{j})$ is an external field. 
Introducing the expectation value $\langle \cdot \rangle_k$, where the average is taken with respect to the action $-S_k[\Psi] + \int_{t,\rr}\Psi^\text{t} J$, it is straightforward to check that the second derivative $W_k^{(2)}$ of \(W_k\) with respect to \(J\)  is given by~\cite{Berges2002}
\begin{equation}
\label{eq:W-properties}
  W_k^{(2)}= \langle \Psi \Psi^\text{t} \rangle_k - \langle \Psi \rangle_k \langle \Psi^\text{t} \rangle_k = G_k,
\end{equation}
where \(G_k\) denotes the $k$-dependent two-point function,  whose entries are given by the \(k\) and \(J\)-dependent version of the response \eqref{defR} and the correlation \eqref{defC} functions. 
Finally, the \(k\)-dependent effective action $\Gamma_k[\Phi]$ is defined as the modified Legendre transform of \(W_k[\Psi]\), given by
\begin{equation}
\label{eq:Gamma-definition}
\Gamma_k[\Phi] = -W_k[J] + \int_{t,\rr} J^\text{t} \Phi - \Delta S_k[\Phi],
\end{equation}
where $J$ is fixed by the condition
\begin{equation}
\Phi = W_k^{(1)} = \langle \Psi \rangle_k.
\end{equation}
The definition of $\Gamma_k[\Phi]$ in Eq.~\eqref{eq:Gamma-definition} is such that~\cite{Berges2002} $\Gamma_{k=\Lambda}[\Phi] \approx S[\Phi]$, i.e., when $k$ is equal to the ultraviolet cutoff $\Lambda$ of the theory, the effective action $\Gamma_k$ reduces to the ``microscopic'' action $S[\Phi]$ evaluated on the expectation value $\Phi$. 
The following relationship can then be derived~\cite{Berges2002}: 
\begin{equation}
\label{eq:Gamma-properties-2}
\Gamma_k^{(1)} = J - \sigma R_k\Phi.
\end{equation}

\section{fRG flow equations}
\label{sec:APP2}
In this appendix we first present the detailed calculation of \(\Delta\Gamma_{n,k}\) Eq.~\eqref{eq:convolution} with \(n=1\) and \(2\). Then, the fRG flow equations for the corresponding couplings \(g_{2n,k}\) which appear in Eq.~\eqref{eq:AnsatzU} are obtained explicitly for the case of vanishing background fields. Details concerning the non-vanishing background Ansatz are then discussed. Finally, we focus on the case \(n>2\), providing analytical formula for the time-dependent part of the fRG flow equations for the various couplings.
\subsection{Calculation of \(G_{0,k}\) and \(\Sigma_k\)}
\label{subsec:app1-1}

Here we want to determine \(G_{0,k}\) in Eq.~\eqref{preDyson}, defined as \(G_k\) in Eq.~\eqref{eq:Gamma-properties-2}, but after replacing \(\Gamma_k^{(2)}\) with its field-independent part. In order to do so, the matrix structure of the second variation $\Gamma^{(2)}_k$ of the modified effective action $\Gamma_k$ is set to  
\begin{equation}
\label{eq:Gamma-second-variation}
\Gamma_k^{(2)}(x,x')= 
\begin{pmatrix}
\displaystyle{\frac{\delta^2\Gamma_k}{\delta\phi(x)\delta\phi(x')}} & \displaystyle{\frac{\delta^2\Gamma_k}{\delta\res(x)\delta\phi(x')}} \\[4mm]
\displaystyle{\frac{\delta^2\Gamma_k}{\delta\phi(x)\delta\res(x')}} & \displaystyle{\frac{\delta^2\Gamma_k}{\delta\res(x)\delta\res(x')}}
\end{pmatrix}.
\end{equation} 
In order to implement the LPA truncation, $\Gamma^{(2)}_k$ above has to be computed on the basis of  the expression for \(\Gamma_k\) given in Eq.~\eqref{eq:effective-action}. Moreover, in order to separate the field-independent  contribution \(G_{0,k}\)  from the field-dependent one  \(\Sigma_k\), we proceed as explained in what follows.
We first set

\begin{equation}
\label{G0}
G^{-1}_{0,k}(q,t,t') = \begin{pmatrix} 0, & \displaystyle{-Z_k\partial_t +\omega_{k,q}} \\ \displaystyle{Z_k\partial_t + \omega_{k,q}}, & D_k 
\end{pmatrix},
\end{equation}
where  the dispersion relation \(\omega_{k,q}\) is given by
\begin{equation}\label{modified-dispersion}
\omega_{k,q}\equiv K_k q^2 + m_k + R_k(q^2),
\end{equation}
with 
\begin{equation}\label{m}
m_k = \displaystyle{\frac{\partial^2 \mathcal{U}_k}{\partial \phi^2}  }(\phi = \bar{\phi}_k).
\end{equation}
Accordingly, we define the field-dependent part \(\Sigma_k\) (we assume $t_0 = 0$ for simplicity), as
\begin{equation}
\label{Sigma0}
\Sigma_k = - \vartheta(t) \begin{pmatrix}
 \displaystyle{\tilde \phi \frac{\partial^3 \mathcal{U}_k}{\partial \phi^3}}, & \displaystyle{\frac{\partial^2 \mathcal{U}_k}{\partial \phi^2} - m_k } \\
  \displaystyle{\frac{\partial^2 \mathcal{U}_k}{\partial \phi^2} - m_k}, & 0
\end{pmatrix}.
\end{equation}
The condition that \(\phi=\bar{\phi}_k\) in the definition of \(m_k\) in Eq.~\eqref{m} is chosen in order to expand the remaining field-dependent part \(\Sigma_k\) as a power series of \((\phi^2-\bar{\phi}^2_k)^2\). This allows one to use a truncation procedure similar to that in the simple case of vanishing background field discussed at the end of Sec.~\ref{subsec:fRG}, as we discuss also below in  App.~\ref{subsec:app1-5}.

After a Fourier transform in space, \(G_{0,k}\) is obtained from Eq.~\eqref{G0}:
\begin{equation}
\label{eq:g0k}
G_{0,k}(q,t,t') = \begin{pmatrix}
\mathcal{C}_{0,k}(q,t,t') & \mathcal{R}_{0,k}(q,t,t')
\\ \mathcal{R}_{0,k}(q,t',t) & 0
\end{pmatrix},
\end{equation}
where
\begin{subequations}
\label{respcorr}
\begin{align}
& \label{R0} \mathcal{R}_{0,k}(q,t,t') = \vartheta(t-t') \frac{1}{Z_k }  \ee^{- \frac{\omega_{k,q}}{Z_k}(t-t')},\\
&  \label{C0} \mathcal{C}_{0,k}(q,t,t')  = 2 D_k \int_{t_0}^{+\infty} \mathcal{R}_{0,k}(q,t,s)\mathcal{R}_{0,k}(q,t',s).
\end{align}
\end{subequations}
We note that in Eq.~\eqref{eq:g0k} the vanishing  entry is due to the fact that the field-independent part of  the second derivative of \(\Gamma\) with respect to \(\phi\) evaluated in the minimum configuration vanishes, as a consequence of causality~\cite{Tauberbook2014}. Moreover, the correlation function  in Eq.~\eqref{respcorr} respects the Dirichlet condition, which is enforced in our Ansatz, as described near Eq.~\eqref{eq:effective-action}.

\subsection{Calculation of \(\Delta\Gamma_{1,k}\) and \(\Delta\Gamma_{2,k}\)}
\label{subsec:app1-2}
In order to calculate the \(\Delta\Gamma_{n,k}\)'s terms in Eq.~\eqref{eq:convolution} we use the so-called optimized regulator~\cite{Litim2000} \(R_k(q)\), given by 
 \begin{equation}
\label{Litim}
R_k(q)=Z_k \vartheta(k^2-q^2)(k^2-q^2).
\end{equation}
First, note that $R_k(q)$ fulfills the conditions in Eqs.~\eqref{eq:cutoff}, introduced in order to construct the fRG equation.
 By using the optimized cutoff function above it is possible to compute  analytically the relevant  integrals appearing in the \(\Delta\Gamma_{n,k}\)'s terms. This is due to two simplifications:   first,  the \(k\)-derivative of the cutoff function \(R_k\) in Eq.~\eqref{eq:convolution} is given by a theta function which vanishes for \(q^2 > k\), i.e.,  \(k \partial_k R_k \propto \vartheta(k^2-q^2)\).  This implements a modified ultraviolet cutoff in the \(q\) integrals in Eqs.~\eqref{eq:convolution}.  The second simplification, which is a consequence of the first, is due to the fact that the modified dispersion relation \(\omega_{k,q}\) in Eq.~\eqref{modified-dispersion} simplifies to \(\omega_{k} = \omega_{k,k=q} = K_k k^2 + m_k\) in the \(q\) interval allowed by \(\partial_k R_k(q)\). 
 The regulator $R_k(q^2)$ thus implements an infrared cutoff \(\sim k^2\), rendering the dispersion $\omega_{k,q}$ indipendent of $q$ for $q\leq k$. 
 Accordingly, the two-time function \(G_{0,k}(q,t,t')\) in Eq.~\eqref{eq:g0k} becomes independent of the momentum \(q\), and will be simply indicated by \(G_{0,k}(t,t')\). 
 
 The first term that we need to calculate is \(\Delta\Gamma_{1,k}\), given by Eq.~\eqref{eq:convolution} with \(n=1\).
 Taking advantage of these two simplifications, the \(q\) dependence, contained in the \(k\)-derivative of the cutoff function, can be factorized and  the following expression for \(\Delta\Gamma_{1,k}\) is obtained: 
 \begin{subequations}\label{d1}
\begin{align}
\Delta \Gamma_{1,k} &=\label{D0} \frac{1}{2} \int_q  \nonumber { \partial_k R_k(q) } 
\\ & \quad \times  \int_{t, t_1} \text{tr}  \Big[ G_{0,k}(t,t_1)\Sigma_k(t_1) G_{0,k}(t_1,t) \sigma \Big], 
\\ & = \label{D1} K_k k^{d+1} \nonumber \frac{a_d}{d} \left(1-\frac{\eta}{d+2} \right) \ 
\\ & \quad \times  \int_{0}^\infty  dt_1 \   \ \Sigma_{11,k}(t_1) \mathcal{F}_{1,k}(t_1),
\end{align}  
\end{subequations}
 where \(\Sigma_{11,k}\) is the \((1,1)\) entry of the matrix \(\Sigma_k\) defined in Eq.~\eqref{Sigma0}, while the time-dependent function \(\mathcal{F}_{1,k}(t_1)\) is defined as 
\begin{equation}
\label{f1}
\mathcal{F}_{1,k}(t_1)\equiv \int_0^{t_1} dt \   \text{tr}  \Big[ G_{0,k}(t,t_1) \sigma_2 G_{0,k}(t_1,t) \sigma \Big],
\end{equation}
with \(\sigma_2 =\left( \begin{smallmatrix}
1 & 0 \\ 0 & 0 
\end{smallmatrix}\right)\). 
The non-diagonal terms of $\Sigma_k$ do not appear in the final result \eqref{D1}, as required by causality \cite{ Tauberbook2014}, since they would be multiplied by a factor $\vartheta(t-t')\vartheta(t'-t) = 0$.
Then, by computing the trace  in Eq.~\eqref{f1} one finds
\begin{subequations}
\begin{align}
\mathcal{F}_{1,k}(t_1) & =   2 \int_0^{t_1} dt \  \mathcal{C}_{0,k}(t_1,t)\mathcal{R}_{0,k}(t_1,t)
\\ & =  \label{F33} 4 D_k \int_0^{t}ds  \ \mathcal{R}_{0,k}^2(t_1,s)(t_1-s)
\\& = \label{F3} \frac{D_k}{Z_k \omega_k^2} \left[1-e^{-2 \frac{\omega_k t}{Z_k} } f_{1}(\omega_k t_1)  \right],
\end{align}
\end{subequations}
 where, in order to obtain Eq.~\eqref{F33}, we have used the expression of \(\mathcal{C}_{0,k}\) in Eq.~\eqref{C0} together with the LPA property, which follows from Eq.~\eqref{R0}, that \(\mathcal{R}_{0,k}(t_1,t)\mathcal{R}_{0,k}(t,s) = \mathcal{R}_{0,k}(t_1,s)\vartheta(t_1-t)\vartheta(t-s)\), and we have performed the integral over \(t\); in order to obtain Eq.~\eqref{F3}, the time integral over \(s\) has been computed by using the analytical expression of \(\mathcal{R}_{0,k}\) given in Eq.~\eqref{R0}. Finally, the expression of \(\Delta\Gamma_{1,k}\), where the Ansatz for \(\mathcal{U}_k\) is still unspecified, is given by Eq.~\eqref{D1}, together with Eq.~\eqref{F3}.

The term \(\Delta\Gamma_{2,k}\) is more complicated than the previous one since, unlike \(\Delta\Gamma_{1,k}\), it involves fields evaluated at different time coordinates. In fact, following the steps with which we have obtained Eq.~\eqref{D1}, but for the case \(n=2\) in Eq.~\eqref{eq:convolution}, one obtains
\begin{equation}
\begin{split}
\label{eq:delta2}
\Delta \Gamma_{2,k} & \propto K_k k^{d+1} \frac{a_d}{d}  \left( 1-\frac{\eta}{d+2}\right)  \\ & \quad \times \int_{t_1,t_2} 2 \Sigma_{11,k}(t_1) \Sigma_{12,k}(t_2) \mathcal{F}_{2,k}(t_1,t_2),
\end{split}
\end{equation}
where the two-time-dependent function \(\mathcal{F}_{2,k}(t_1,t_2)\) is given by 
\begin{equation}
\label{deff3}
\begin{split}
 \mathcal{F}_{2,k}(t_1,t_2)   &= 
\int_0^{t_1} dt  \   \text{tr}  \left[ G_{0,k}(t,t_1)\sigma_2  \right.     
\\ & \quad  \quad  \quad \times\left. G_{0,k}(t_1,t_2)\sigma G_{0,k}(t_2,t)\sigma \right] .
\end{split}
\end{equation}
In Eq.~\eqref{eq:delta2}  we have retained only the contribution  proportional to \(\tilde\phi\), i.e., proportional to \(\Sigma_{11,k}\) (see Eq.~\eqref{Sigma0}), since they are the only ones which renormalize the couplings \(g_{2n,k}\), as one can see from their definition in Eq.~\eqref{eq:vertexp-maintext}. Other contributions, which renormalize the noise term \(D_k\), are discussed in Ref.~\onlinecite{Chiocchetta2016}.
As one can see from Eq.~\eqref{eq:delta2}, the kernel \(\mathcal{F}_{2,k}(t_1,t_2)\) is a two-time function from which one has to extract  its local part, accordingly to the LPA Ansatz \eqref{eq:effective-action},  
in order to obtain the contribution of the  \(\Delta\Gamma_{2,k}\)'s  proportional to the terms which contain the coupling terms. This can be achieved by substituting  
\(
\phi(t_2) \rightarrow \phi(t_1)
\) in Eq.~\eqref{eq:delta2} \cite{Chiocchetta2016}. Other contributions, which renormalize the derivative term \(Z_k\), are discussed in Ref.~\onlinecite{Chiocchetta2016}. If the previous substitution is made, Eq.~\eqref{eq:delta2} becomes  
\begin{equation}
\label{d2}
\begin{split}
\Delta \Gamma^l_{2,k}  &\propto K_k k^{d+1}\frac{a_d}{d} \left( 1-\frac{\eta}{d+2}\right) 
\\ & \quad  \int_{0}^\infty dt_1 \ 2 \Sigma_{11,k}(t_1) \Sigma_{12,k}(t_1) \mathcal{F}^l_{2,k}(t_1),
\end{split}
\end{equation}
where  the superscript \(l\) denotes the local part  of \(\mathcal{F}_{2,k}\) defined in Eq.~\eqref{deff3}, i.e.,
\begin{subequations}
\label{f3}
\begin{align}
\mathcal{F}^l_{2,k}(t_1) & = \int_0^{t_1}  dt_2 \ \mathcal{F}_{2,k}(t_1,t_2)  \\ & = \label{f43} 4 D_k \int_0^{t_1} ds \  \mathcal{R}_{0,k}^2(t_1,s) (t_1-s)^2 \\ &  =  \label{f4}  \frac{D_k}{Z_k \omega_k^3} \left[1-e^{-2 \frac{\omega_k t}{Z_k} } f_{2}(\omega_k t_1) \right].
\end{align}
\end{subequations}
In order to obtain Eq.~\eqref{f43} we have computed the trace in Eq.~\eqref{deff3} and then the intermediate time integrals over \(t\) and \(t_2\), similarly to what has been done in the computation of Eq.~\eqref{f1} leading to Eq.~\eqref{F3}. In order to obtain Eq.~\eqref{f4}, we computed the  time integral over \(s\) in Eq.~\eqref{f43} by using the analytic expression of \(\mathcal{R}_{0,k}\) in Eq.~\eqref{R0}. Finally, the local part \(\Delta\Gamma^l_{2,k}\) of \(\Delta\Gamma_{2,k}\), proportional to \(\tilde\phi\), is given by Eq.~\eqref{d2} together with \(\mathcal{F}^l_{2,k}(t_1)\) in Eq.~\eqref{f4}.
We have thus shown explicitly how the extraction of the local contribution in \(\Delta\Gamma_{2,k}\) is implemented in the simple case of \(\bar{\phi}_k=0\).

\subsection{Case with \(\bar{\phi}=0\) and \(n_\text{tr}=2\)}
\label{subsec:app1.3}
Here we want to derive the fRG flow equation for the parameter \(r_k\) and \(g_k\) introduced in the Ansatz for the effective potential \(\mathcal{U}_k\) given by Eq.~\eqref{Uphi0}, for which \(K_k=D_k=Z_k=1\), as discussed in Sec.~\ref{sec:truncation-symmetric}. 
The field-dependent part \(\Sigma_k\) of \(\Gamma^{(2)}\) is given by Eq.~\eqref{sigmaf0}.
Then, by taking advantage of Eq.~\eqref{D1}, with Eqs.~\eqref{d2} and \eqref{sigmaf0}, one obtains 
\begin{equation}
\label{GF0}
\begin{split}
  \frac{d \Gamma_k}{dk} \propto &   \left( k^{d+1}  \frac{a_d}{d} \right) \ 
 \\ &  \times  \int_{0}^\infty  dt_1  \tilde \phi(t_1) \left[   g_k  \  \phi(t_1) \mathcal{F}_{1,k}(t_1) \right. 
\\ &  \quad \quad + \left. g^2_k     \  \phi^3(t_1) \mathcal{F}^l_{2,k}(t_1) \right],
\end{split}
\end{equation}
where we have considered only linear contributions  in \(\tilde\phi\), in order to obtain the fRG flow equations of the coupling terms  \(g_{2n,k}\) in Eq.~\eqref{eq:vertexp-maintext}, given by \(r_k\) and \(g_k\). The time-dependent functions \(\mathcal{F}_{1,k}\) and \(\mathcal{F}_{2,k}^l\) are the same as those obtained previously, given  by Eqs.~\eqref{F3} and \eqref{f4} respectively, with the dispersion relation \(\omega_k = k^2 + r_k\), with \(m_k = r_k\).

Finally, the fRG equation for the coupling \(r_k\) and \(g_k\) in  Eqs.~\eqref{eq:beta}, are obtained by means of Eqs.~\eqref{flowcoupling} with \(\frac{d \bar{\phi}_k}{dk}=0\) and the \(k\)-derivative of \(\Gamma_k\) given by Eq.~\eqref{GF0}, setting \(n=1\) and \(n=2\), respectively.

\subsection{Case with \(\bar{\phi}\neq 0\) and \(n_\text{tr} = 2\)}
\label{subsec:app1-5}

The construction of the fRG equations for the  couplings  \(g_{2n,k}\) which define the Ansatz for the effective potential \(\mathcal{U}_k\), given in Eq.~\eqref{potentialordered}, is discussed here  for the case of non-vanishing background analyzed in Sec.~\ref{sec:truncation-ordered}. 
According to the definition of \(\Sigma_k\) in Eq.~\eqref{Sigma0}, one finds
\begin{equation}
\label{eq:self-energy-ordered}
\Sigma_k(x) = -g_k\,  \vartheta(t)
\begin{pmatrix}
\widetilde{\rho}(x)		& \rho(x)-\bar{\rho}_k \\
\rho(x)-\bar{\rho}_k 	&	 0
\end{pmatrix},
\end{equation}
where we define
\begin{equation}
\label{eq:invariants}
\rho\equiv\frac{\phi^2}{2}, \qquad \widetilde{\rho}\equiv \res\phi, \ \ \ \text{ and} \qquad \bar{\rho}_k \equiv \frac{\bar{\phi}_k^2}{2}, 
\end{equation}
while $G_{0,k}$ is defined according to Eq.~\eqref{eq:g0k} with  \(m_k\) as in Eq.~\eqref{m} given by
\begin{equation}
\label{eq:effective-mass}
m_k=\frac{2}{3}\bar{\rho}_k g_k,
\end{equation}
ensuing from definition~\eqref{m}.
The use of the $\mathbb{Z}_2$ invariants $\rho$ and $\widetilde{\rho}$ introduced in  Eq.~\eqref{eq:invariants} is customary in the context of fRG~\cite{Delamotte2007} and it helps in simplifying the notation in what follows. 
The form of $\Sigma_k(x)$ in Eq.~\eqref{eq:self-energy-ordered} allows us to express the r.h.s.~of the fRG equation~\eqref{eq:Wett2} as a power series of $\rho-\bar{\rho}_k$: this is done in the spirit of the discussion below Eq.~\eqref{sigmaf0}. Accordingly, together with the vertex expansion \eqref{eq:vertexp-maintext},  this provides a way to unambiguously identify the renormalization of the terms appearing in the potential $\mathcal{U}_k$ in Eq.~\eqref{potentialordered}. In fact, $\bar{\rho}_k$ and the coupling $g_k$ are identified as~\cite{Delamotte2007}
\begin{equation}
\label{eq:projections-rules-1}
\frac{d \mathcal{U}_k}{d \rho}\biggr|_{\rho = \bar{\rho}_k} \!\!\!\!\! = 0, \quad 
\frac{g_k}{3} = \frac{d^2 \mathcal{U}_k}{d \rho^2}\biggr|_{\rho=\bar{\rho}_k}, 
\end{equation}
where the first condition actually defines $\bar{\rho}_k$ as the minimum of the potential. In terms of the effective action $\Gamma_k$, Eqs.~\eqref{eq:projections-rules-1} become
\begin{equation}
\label{eq:projections-rules-2}
\frac{\delta \Gamma_k}{\delta \widetilde{\rho}}\biggr|_{\substack{\widetilde{\rho}=0\\ \rho=\bar{\rho}_k}} \!\!\!\!\!= 0, \text{ and} \quad 
\frac{g_k}{3} = \frac{\delta^2 \Gamma_k }{\delta\widetilde{\rho}\,  \delta\rho}\biggr|_{\substack{\widetilde{\rho}=0\\ \rho=\bar{\rho}_k}}.
\end{equation}
By taking a total derivative with respect to $k$ of each equality in Eqs.~\eqref{eq:projections-rules-2}, one finds
\begin{align}
& \frac{\delta }{\delta \widetilde{\rho}} \frac{d \Gamma_k}{d k}\biggr|_{\substack{\widetilde{\rho}=0\\ \rho=\bar{\rho}_k}} + \frac{\delta^2 \Gamma_k}{\delta \widetilde{\rho}\,\delta \rho}\biggr|_{\substack{\widetilde{\rho}=0\\ \rho=\bar{\rho}_k}} \frac{\dd \bar{\rho}_k}{\dd k}=0, \label{eq:rho0-flow-general}\\ 
& \frac{1}{3} \frac{\dd g_k}{\dd k} = \frac{\delta^2 }{\delta \widetilde{\rho}\,\delta \rho} \frac{d \Gamma_k}{d k}\biggr|_{\substack{\widetilde{\rho}=0\\ \rho=\bar{\rho}_k}} + \frac{\delta^3 \Gamma_k}{\delta \widetilde{\rho}\,\delta^2 \rho}\biggr|_{\substack{\widetilde{\rho}=0\\ \rho=\bar{\rho}_k}} \frac{\dd \bar{\rho}_k}{\dd k},\label{eq:g-flow-general}
\end{align}
which, after replacing $d \Gamma_k /d k$ with the fRG equation~\eqref{eq:wetterich}, render the flow equations for $\bar{\rho}_k$ and $g_k$. For the case of the potential $\mathcal{U}_k$ in Eq.~\eqref{potentialordered}, by using Eq.~\eqref{eq:projections-rules-2}, the set of flow equations~\eqref{eq:rho0-flow-general} and~\eqref{eq:g-flow-general} simplifies as
\begin{align}
\frac{\dd \bar{\rho}_k}{\dd k} &= - \frac{3}{g_k}\frac{\delta }{\delta \widetilde{\rho}} \frac{\dd \Gamma_k}{\dd k}\biggr|_{\substack{\widetilde{\rho}=0\\ \rho=\bar{\rho}_k}} = - \frac{3}{g_k}\frac{\delta \Delta \Gamma_{1,k}}{\delta \widetilde{\rho}}  \biggr|_{\substack{\widetilde{\rho}=0\\ \rho=\bar{\rho}_k}}, \label{eq:rho0-flow-ordered}\\
\frac{1}{3} \frac{\dd g_k}{\dd k} & = \frac{\delta^2 }{\delta \widetilde{\rho}\,\delta \rho} \frac{\dd \Gamma_k}{\dd k}\biggr|_{\substack{\widetilde{\rho}=0\\ \rho=\bar{\rho}_k}} = \frac{\delta^2\Delta\Gamma_{2,k} }{\delta \widetilde{\rho}\,\delta \rho} \biggr|_{\substack{\widetilde{\rho}=0\\ \rho=\bar{\rho}_k}}\label{eq:g-flow-ordered},
\end{align}
where we used Eq.~\eqref{eq:Wett2} with $\Delta \Gamma_{1,k}$ and $\Delta \Gamma_{2,k}$ defined in Eqs.~\eqref{eq:convolution} in terms of the $\Sigma_k(x)$ in Eq.~\eqref{eq:self-energy-ordered}.  
The explicit form of the flow equations comes from a calculation analogous to the one discussed in Sec.~\ref{sec:truncation-symmetric} and then detailed in App.~\ref{subsec:app1-2} (see Eqs.~\eqref{d1} and~\eqref{d2}). In particular, the flow of $m_k$, defined in Eq.~\eqref{eq:effective-mass}, takes contributions from  the flow equations for both $\bar{\rho}_k$ and $g_k$.

Then, it is possible to construct the fRG flow equations for the dimensionless couplings
\begin{equation}
\label{eq:couplings-dimensionless}
\widetilde{m}_k=\frac{2}{3}\frac{\bar{\rho}_k g_k}{K_k k^2}, \ \ \text{ and}\quad \widetilde{g}_k=\frac{a_d}{d}\frac{D_k}{Z_kK_k^2}\frac{g_k}{k^{4-d}},
\end{equation}
analogous to Eqs.~\eqref{eq:fixed point}.
Considering the solutions of the fRG equations with vanishing \(k\)-derivative of these dimensionless parameters one obtains equations similar to Eqs.~\eqref{eq:fixed point} that finally allows the determination of the fixed point values. In order to solve these fRG equations, one has to  supplement them with those  involving  the anomalous dimension \(\eta\) \eqref{anomalousdimension}, as one can see from the expression of the \(\Delta\Gamma_{n,k}\)'s terms computed in Eqs.~\eqref{D1} and \eqref{d2}. 
\subsection{Higher-order \(\Delta\Gamma_{n,k}\)'s terms}
\label{subsec:app1-4}
We discuss here what happens in the general case, i.e., with \(n_\text{tr}\geq 2 \). Correspondingly,  all the \(\Delta\Gamma_{k,n}\)'s terms with \(n\) up to \(n_\text{tr}\) have to be computed. For these terms the associated time-dependent function \(\mathcal{F}_{n,k}\), that appear in Eqs.~\eqref{d1} and \eqref{eq:delta2}  for \(n=1\) and \(n=2\) respectively,  depend upon \(i=1,\dots,n\) different times. The procedure for extracting the local contributions, with which we have obtained the last equation in Eq.~\eqref{d2}, has to be extended to all the times \(t_i\) other than \(t_1\). In order to retain linear contribution in \(\tilde\phi\), as before, it suffices to replace \(\phi(t_i) \rightarrow \phi(t_1)\).
In fact, one can show that the structure of  Eqs.~\eqref{F3} and \eqref{f4} is robust for higher-order terms (\(n>2\)) and all of them leads to terms of the form
\begin{equation}
\label{Fl}
\mathcal{F}^l_{n,k}(t_1)=  \frac{D_k}{Z_k \omega_k^{n+1}} \left[1-e^{-2{\omega_k t}/{Z_k} } f_n\left(\frac{\omega_k}{Z_k} t\right ) \right],
\end{equation}
 with \(f_n(x)\) being a polynomial in \(x\). Accordingly,  the exponential decay over time of the fRG flow equations \eqref{eq:beta} is a general feature within LPA approximation. Because of this, one can recover the long-time equilibrium fRG equations. 
This last equation is the proof of the statement done in the main text about the fact that any time-dependence in the fRG flow equation is of the exponentially decaying form as the one in Eqs.~\eqref{eq:beta}; accordingly, the equilibrium flow equations are retrieved in the long-time limit, since \(\eta_D = \eta_Z\), as discussed around Eq.~\eqref{anomalousdimension} in the main text.

In addition, by inspection of the time-dependent part of the \(\Delta\Gamma_{n,k}\)'s terms and based on  previous considerations, an analytical formula  for any localized kernel \(\mathcal{F}^l_{n,k}(t_1)\) can be obtained as
\begin{subequations}
\label{FL}
\begin{align}
\mathcal{F}^l_{n,k}(t_1) &= \label{general1} \frac{2^{n+1}}{n!}D_k \int_0^{t_1}ds \ \mathcal{R}_{0,k}^2(t_1,s)(t_1-s)^n \\ & = \label{general2} \frac{2^{n+1}}{n!}D_k \frac{\dd^n}{\dd (-2 \omega_k/Z_k)^n} \mathcal{F}\left(\frac{\omega_k}{Z_k} t\right),
\end{align}
\end{subequations}
where in the second equality \(\mathcal{F}\) is defined as
\begin{subequations}
\label{FFL}
\begin{align}
\mathcal{F}\left(\frac{\omega_k}{Z_k} t_1\right) &=\label{generalpenultima} \int_0^{t_1} dt \  \mathcal{R}_{0,k}^2(t_1,t)
\\ & =\label{generalultima} \frac{1}{2 Z_k  \omega_k} \left(1-e^{-2 {\omega_k t_1}/{Z_k} } \right),
\end{align}
\end{subequations}
and we used the analytical expression of  \(\mathcal{R}_{0,k}\) given by Eq.~\eqref{R0}.
We note that Eq.~\eqref{general1} reproduces Eqs.~\eqref{F33} and \eqref{f43} respectively for \(n=1\) and \(n=2\). Finally, Eq.~\eqref{generalultima} is simply obtained  by computing the  integral over \(t\) in Eq.~\eqref{generalpenultima} with \(\mathcal{R}_{0,k}\) given in Eq.~\eqref{R0}.

The formulas in Eqs.~\eqref{general2} and \eqref{generalultima} are very useful in order to implement analytically the LPA approach discussed here for the case of higher-order  terms (\(n_\text{tr}> 2\)), thus calculating the corresponding \(f_{n,k}(x)\) from the comparison with Eq.~\eqref{Fl}. Taking advantage of these analytical formulas a code in Mathematica has been used to compute the \(\Delta\Gamma_{k,n}\)'s terms for \(n\) up to \(n = n_\text{tr}^\text{max}=6\), as explained in Sec.~\ref{sec:results}.

\section{ Long-time aging regime }
\label{C}
In this appendix, we first detail how in the absence of the background field the aging regime for the response and the correlation functions, given by Eqs.~\eqref{scalingLPA}, is obtained. Then we consider the effect of the presence of a background field, detailing the way in which we obtain the expression for the reduced parameter \(r_k/Z_k\), needed  for solving Eqs.~\eqref{eq:motionlpa'}. Finally, we compare our results for \(\theta\) with those obtained in Ref.~\onlinecite{Chiocchetta2016}, proving the equivalence of these two approaches.

\subsection{Case  \(\bar{\phi}=0\)}
\label{sec:proof}
Here we detail the calculations that led to the predictions  of the response and the correlation functions in the aging regime, given by Eq.~\eqref{scalingLPA}, for the case of vanishing background field with \(n_\text{tr}=2\), starting from Eqs.~\eqref{eq:greens-motion-lpa} and \(r\) given by Eq.~\eqref{raging}.
The response function in Eqs.~\eqref{scalingLPA} is readily obtained in the long-time limit of the aging limit, solving Eq.~\eqref{Rreduced} with \(r(t)\) given by Eq.~\eqref{raging} with \(t>t'\gg \Lambda^{-2}\). For the correlation function it is less straightforward to extract the behavior in the aging regime: one can rewrite Eq.~\eqref{Creduced} as 
\begin{equation}
\label{correlation}
\mathcal{C}(t,t') = 2 \mathcal{R}(t,t') \int_0^{t'} ds \  \mathcal{R}^2(t',s),
\end{equation}
where we used the identity
\(\mathcal{R}(t,s)=\mathcal{R}(t,t')\mathcal{R}(t',s)\), which follows from the LPA approximation~\eqref{Rreduced}.
Accordingly, the integral in the equation above should be computed in the regime \(t'\gg \Lambda^{-2}\). 
This amount of studying 
\begin{equation}\label{deflim}
\lim_{t'\rightarrow \infty} \int_0^{t'}ds \mathcal{R}^2(t',s) , 
\end{equation}
where
\begin{equation}\label{genR}
\mathcal{R}(t',s) = \exp\left(-\int_{s}^{t'}ds' r^*(s')\right).
\end{equation}

The LPA prediction for the parameter \(r(t)\) in  the case of a critical quench of the model is given by Eq.~\eqref{r}, that we report here   
\begin{equation}
\label{RAGING}
\begin{split}
r^*(t) &= - \frac{\theta}{t}\left[1-\tilde{F}_r(\Lambda^z t) \right],
\end{split}
\end{equation}
where \(\tilde F_r\) is a function which decays exponentially fast as \(t\) grows. 
Let us simplify the integral in Eq.~\eqref{deflim}, taking advantage of Eqs.~\eqref{genR} and \eqref{RAGING}:
\begin{equation}\label{A3}
\begin{split}
& \int_0^{t'} ds \ \mathcal{R}^2(t',s)  
\\  = &  t' \int_0^1 d\tau\   \tau^{-2\theta} \text{exp}\left[-2\theta \int_{\tau}^1 \frac{d\tau'}{\tau'} \tilde{F}_r(\Lambda^z t' \tau' ) \right],
\end{split}
\end{equation}
where, in the last equality, we made the following changes of variable: \(s =t' \tau\) and \(s'= t' \tau' \).

We recall that the  aging regime is reached for \(A = \Lambda^z t' \rightarrow \infty\).
We break the integral in \(\tau\) which appears in  Eq.~\eqref{A3} as \(\int_0^1\dd\tau =  \int_{A^{-1}}^1\dd\tau+ \int_0^{A^{-1}}\dd\tau \).
The integral  $\int_{A^{-1}}^1d\tau$ is given, in the aging limit, by $\int_0^1 d\tau \ \tau^{-2\theta} = (1-2\theta)^{-1}$, where we assumed that $\theta <1/2$.
In the following, we prove that the remaining integral over $\int_0^{A^{-1}}d\tau$ gives a vanishing contribution.
To do so, we break the  integral  in \(\tau'\) which appear in  Eq.~\eqref{A3}, as \(\int_{\tau}^1 d\tau' = \int_{\tau}^{A^{-1}}d\tau'+\int_{A^{-1}}^1d \tau'\). Since the integral over $\int_{A^{-1}}^1d\tau'$ converges, it gives an overall vanishing contribution when integrated over $\int_0^{A^{-1}}d\tau$ in the aging regime. Let us now focus on the remaining integral given by
\begin{equation}
\begin{split}& \int_0^{A^{-1}} d\tau \ \tau^{-2\theta} \text{exp}\left[-2\theta \int_\tau^{A^{-1}}\frac{d\tau'}{\tau'} f_r(A\tau' ) \right].
\end{split}
\end{equation}
 With the change of variable \(u = A \tau \) we obtain
\begin{equation}
A^{-1+2\theta} \int_0^1 du\ u^{-2\theta} \text{exp}\left[-2\theta \int_{u/A}^{1/A} \frac{d\tau'}{\tau'} f_r(A \tau') \right],
\end{equation}
and, as long as \(-1+2\theta < 0 \), this integral gives a vanishing contribution in the aging limit \(A  = \Lambda^z s \rightarrow \infty\) to the limit expression given by Eq.~\eqref{deflim}.

Summarizing, we have obtained 
\begin{equation}
\lim_{t'\rightarrow\infty}\int_0^{t'} ds \ \mathcal{R}^2(t',s) = \frac{t'}{1-2\theta},
\end{equation}
which, when inserted in Eq.~\eqref{correlation}, with Eq.~\eqref{RAGING}, gives Eq.~\eqref{scalingLPA}, as anticipated in the main text.

\subsection{Case  \(\bar{\phi}\neq 0 \)}
\label{sec:Apptheta}

Here we discuss how, in the presence of a background field, one can solve the equations \eqref{eq:motionlpa'} for the reduced two-time functions.

First, we prove that  equations similar to Eqs.~\eqref{r} and \eqref{raging} can be obtained for the reduced parameter \(r/Z\) which enters  Eq.~\eqref{eq:motionlpa'}.
In order to obtain these equations one can use a simplification that appears at the level of the flow equation for it, i.e.,
\begin{equation}\label{rbn}
\frac{\dd}{\dd k}\frac{r_k}{Z_k} = \frac{1}{Z_k}\frac{\dd r_k}{\dd k} - \eta_K \frac{r_k}{Z_k}.
\end{equation}
In fact, according to our previous analysis, we retain only the explicit time dependence of \(G_{0,k}\) which appears in the \(\Delta\Gamma_{n,k}\)'s terms, as we have discussed in Sec.~\ref{subsec:fRG} near Eqs.~\eqref{eq:beta}. This amounts to the fact that the second term on the r.h.s. in the previous equation is time-independent, thus it will simply renormalize  \(r_\Lambda\) and therefore the value it has to take in order to obtain a vanishing long-time limit of \(r(t)\), as we did  in order to obtain Eq.~\eqref{r} from Eq.~\eqref{eq:r-eq} in Sec.~\ref{subsec:Green}.

 The fRG equation \eqref{rbn}  for the reduced  parameter \(r_k/Z_k\)   is obtained, e.g., for the case of the non-vanishing background field Ansatz with \(n_\text{tr}=2\), once the corresponding flow equation for \(\bar{\rho}_k\) and \(g_k\) are derived, as explained in Sec.~\ref{ultima}. 
 Accordingly, the flow equation of \(r_k/Z_k\) is  proportional to field derivatives of \(\Delta\Gamma_{1,k}\) and \(\Delta\Gamma_{2,k}\), given by Eqs.~\eqref{eq:rho0-flow-ordered} and \eqref{eq:g-flow-ordered}. Thus, it is proportional to \(\mathcal{F}_{1,k}\) and \(\mathcal{F}^l_{2,k}\), given explicitly by Eqs.~\eqref{F3} and \eqref{f4}. These two terms are equivalent for the analysis that follows, since their time dependence is always through \({\omega_k t}/{Z_k} \). 
 Considering only one of these terms, the flow equation for the reduced parameter \(r_k/Z_k\)  given by \begin{equation}
\label{rkansatz}
\frac{1}{Z_k} \frac{d r_k(t)}{dk} = \tilde A_k \ k \frac{K_k}{Z_k} \frac{D_k}{Z_k}  \left[1-  F_k \left(\frac{\omega_k}{Z_k} t\right)\right],
\end{equation}
where \(\tilde A_k\) is given by the dimensionless time-independent part of the corresponding fRG flow equation (\(\tilde A_k = \tilde g_k /(1+\tilde r_k)^2\) in  Eq.~\eqref{eq:r-eq}), and \(F_k\) is an exponentially vanishing function of its argument. In the vicinity of the infrared fixed point, i.e., for  \(k\rightarrow 0\), the factor \({\omega_k t}{Z_k}\) behaves as \(\sim k^z t\), while \(k \frac{K_k}{Z_k}\frac{D_k}{Z_k} \sim k^{z-1} \), as a consequence of Eq.~\eqref{anomalousdimension}.


The critical initial-slip exponent  \(\theta\) is  calculated from the integral over the cutoff \(k\) of Eq.~\eqref{rkansatz} (see the discussion which leads to Eq.~\eqref{thetalpa} and  apply it to the  case of non-vanishing background Ansatz, i.e., replacing \(r_k\) with its reduced version \(r_k/Z_k\), as explained in Sec.~\ref{ultima}). In the vicinity of the infrared fixed point, one finds 
\begin{subequations}
\begin{align}
 -\frac{\theta}{t}  & \sim   \lim_{t\rightarrow \infty}\int_\Lambda^0 dk \ \frac{1}{Z_k} \frac{d r_k^*(t)}{dt},
 \\ & = \label{finaltheta}\lim_{t\rightarrow \infty}\frac{1}{z t} \tilde{A}^* \int_{\Lambda^z t}^0  dx \  \left[-  F^*_k \left( x \right)\right],
 \end{align} 
\end{subequations}
where  \(r^*_k\) means that the initial parameter \(r_\Lambda\) is  tuned in order to have \(r^*(t) \rightarrow 0 \) for \(t \rightarrow \infty\), as explained in the main text around Eq.~\eqref{defmass}, and the substitution \(x = k^z t\) is made in order to obtain Eq.~\eqref{finaltheta}. 
This equation  is the proof of the statement done in the main text about the fact that relations similar  to Eqs.~\eqref{r} and~\eqref{raging} are obtained also in the non-vanishing background field approximation for the reduced parameter \(r/Z\).

\subsection{Comparison with the short-time analysis}
\label{comparison}
Here we compare the predictions derived in Sec.~\ref{sec:5}
for \(\theta\) with those  obtained in Ref.~\onlinecite{Chiocchetta2016}.
There, \(\theta\) was calculated via an analysis of the short-time behavior, i.e., focusing on the limit in which the waiting time $t'$ is approximately the initial time \(t_0\), i.e., \(t'\sim t_0 =0\). 
From general scaling arguments for the aging dynamics~\cite{Calabrese2004} the anomalous dimension \(\eta_0\) of the boundary field \(\tilde\varphi_0\)  is related to \(\theta\) by \cite{Chiocchetta2016}
\begin{equation}
\label{thetashorttime}
\theta = -\eta_0/z.
\end{equation} 

In the case with \(\bar{\phi}_k=0\) and \(n_\text{tr}=2\) given by Eq.~\eqref{Uphi0},  \(z=2\), according  to the fact that in the lowest order of LPA no anomalous dimension arises. Moreover, the anomalous dimension \(\eta_0\) of the boundary field, given in Ref.~\onlinecite{Chiocchetta2016} by Eq.~(46)  with \(\tilde \tau_0 = +\infty\),  matches the one which can be extracted from Eq.~\eqref{thetalpa} by comparing it with Eq.~\eqref{thetashorttime}. Accordingly, the analysis of the short-time behaviour done in Ref.~\onlinecite{Chiocchetta2016} and at long times presented here provide the same prediction for \(\theta\), as it should be, given that the scaling functions \eqref{eq:greens-scaling} are attained in the regime  \(t/t'\gg 1 \), which encompasses both cases.
The equivalence between the two methods for  calculating \(\theta\) is also valid in the non-vanishing background field approximation, as described in what follows.

In the general case, it follows from the comparison of Eq.~\eqref{thetashorttime} with Eq.~\eqref{finaltheta} that 
\begin{equation}
\label{etalong}
\eta_0 = \tilde A^* \lim_{t \rightarrow \infty} \int_{\Lambda^z t}^0  \dd x \  \left[-{F}_k^*(x)\right].
\end{equation}
From the previous idenitification of \(\tilde A_k\) (see below Eq.~\eqref{rkansatz}) and computing the integral over \(x\) from \(\infty\) to \(0\) of \(F^*(x)\) which appear in Eq.~\eqref{etalong}, where \(F^*_k(x)= \ee^{-2 x(1+\tilde r^*)}f_1(x)\) with \(f_1(x)= 1+2(1+\tilde r^*)x \), it follows that \(\eta_0 = \tilde - A^* / (1+\tilde r^*) \); this expression is  exactly Eq.~(46) in Ref.~\onlinecite{Chiocchetta2016} and Eq.~\eqref{thetalpa} obtained here with \(z=2\).

In order to prove the equivalence between the two methods we consider the analytical formula that, in Ref.~\onlinecite{Chiocchetta2016}, has been used in order to calculate  the anomalous dimension \(\eta_0\) of the boundary fields \(\tilde\varphi_0\), which is given by
\begin{equation}
\label{eta0short}
\eta_0 = - \int_0^\infty \dd t \ k \frac{1}{Z_k}\frac{\dd r_k^*(t)}{\dd k},
\end{equation}
where only the time dependent part in Eq.~\eqref{rkansatz}, i.e., \(F_k(x)\) is retained.
Accordingly, Eq.~\eqref{eta0short} matches exactly with Eq.~\eqref{etalong} if the substitution \(x=k^z t \) is made. The equivalence of the two approaches then follows.

\section{Expression of \(\theta\) for  \(\bar{\phi}_k\neq 0\) and \(n_\text{tr} = 3\)}
\label{formula}

Here we provide the correct expression of \(\theta\) for the case of non-vanishing background field with \(n_\text{tr}=3\), fixing a mistake in Ref.~\onlinecite{Chiocchetta2016} (see Eq.~\eqref{massn3} and discussion around it). For completeness, we further provide the details which allow the numerical computation of \(\theta\).
First, the equations for the fixed-point dimensionless couplings  \(\tilde m^*\), \(\tilde g^*\) and \(\tilde \lambda^*\)  defined above Eq.~\eqref{eq:fixed point} are given by (the superscript \(^*\) which henceforth we omit denotes fixed point values), 
\begin{widetext}
\begin{subequations}
\label{flowbulk}
\begin{align}
\label{flowbulk1}
0  &= (-2+\eta )\tilde m  +  \left(1-\frac{\eta }{d+2}\right)  \frac{2\tilde g}{(1+\tilde m)^2} \left[ 1 + \frac{3}{2}\left(\frac{\tilde m  \tilde \lambda }{\tilde g ^{2}}\right)^2  + \frac{3\tilde m }{1+\tilde m }\left(1+\frac{\tilde m  \tilde \lambda }{\tilde g  ^2} \right)^2\right],
\\
\label{flowbulk2}
 0& =  \tilde g \left[ d-4+2\eta   + \left(1-\frac{\eta }{d+2}\right)\frac{6 \tilde g }{(1+\tilde m )^3} \left(1+\frac{\tilde m  \tilde \lambda }{\tilde g ^2} \right)^2\right] +  \left(1-\frac{\eta }{d+2}\right)\frac{\tilde \lambda }{(1+\tilde m )^2}\left(-2+3\frac{\tilde m  \tilde \lambda }{\tilde g ^2} \right),
 \\
 \label{flowbulk3}
0 &= \tilde \lambda  \left[ 2d-6 + 3\eta  + 30\left(1-\frac{\eta }{d+2}\right)\frac{\tilde g }{(1+\tilde m )^3}\left(1+\frac{\tilde m  \tilde \lambda }{\tilde g ^2}\right)\right] -18\left(1-\frac{\eta }{d+2}\right)\frac{\tilde g ^3}{(1+\tilde m )^4}\left(1+\frac{\tilde m  \tilde \lambda }{\tilde g ^2} \right)^3,
\end{align}
\end{subequations}
\end{widetext}
where  \(\eta=\eta_k\) and  the dimensionless couplings \(\tilde m_k\) and \(\tilde g_k\) are defined in Eq.~\eqref{eq:couplings-dimensionless}, while \(\tilde \lambda_k\) has been defined as
\begin{equation}
\tilde \lambda_k = \frac{a_d}{d} \frac{D_k^2}{Z_k^2 K_k^3}\frac{\lambda_k}{5 k^{6-2d}}.
\end{equation}
Note that in Ref.~\onlinecite{Chiocchetta2016}  this definition of \(\tilde\lambda_k\) was used, although the text therein reported a definition without the  factor \(5\) at the denominator. 
Since Eqs.~\eqref{flowbulk} depend upon the anomalous dimension \(\eta_K\), one has to supplement them with the equation for it, given by 
\begin{subequations}
\begin{align}
\eta_K & = \frac{3 \meff \geff}{(1+\meff)^4}\left( 1 + \frac{\meff\leff}{\geff^2} \right)^2, \label{eq:etaK-RG}\\
\eta_Z & = \left( 1 - \frac{\eta_K}{d+2} \right)\frac{9 \meff \geff}{2(1+\meff)^4}\left( 1 + \frac{\meff\leff}{\geff^2} \right)^2,
\label{eq:etaZ-RG}
\end{align}
\end{subequations}
where we have reported also the anomalous dimension related to the derivative parameter \(Z_k\) and the noise term \(D_k\). 
From Eqs.~\eqref{flowbulk}, using  Eq.~\eqref{eq:etaK-RG}, one can calculate numerically the fixed point values (\(\tilde m^*, \tilde g^*, \tilde\lambda^*\)) depending on the spatial dimensionality \(d\), as we have done using  Eqs.~\eqref{eq:fixed point} in the main text. We find numerically (using Wolfram Mathematica 12.3.1) the following fixed point values of the rescaled couplings in $d=3$ (up to the second  significant digit): 
\begin{equation}\label{eq:FPvaluess}
\tilde{m}^*\simeq0.30, \quad\tilde{g}^*\simeq0.26, \ \  \text{and} \quad\tilde{\lambda}^*\simeq0.04.
\end{equation}
The values $\eta^*_{K,Z}$ of the anomalous dimensions  at this fixed point are  found  by replacing directly Eq.~\eqref{eq:FPvaluess} into the expressions~\eqref{eq:etaK-RG} and~\eqref{eq:etaZ-RG}. The dynamical critical exponent \(z\) is given by the second of Eqs.~\eqref{anomalousdimension}. 
In order to compute $\theta$, we use the general scaling relation for \(\theta\), given by Eq.~\eqref{thetashorttime}, and the value of \(z\) obtained via a long-time analysis of the dynamics, one can focus on the anomalous dimension \(\eta_0\) of the boundary field. Following the calculations discussed in Apps.~\ref{sec:Apptheta} and \ref{comparison}  one obtains
\label{eta0}
\begin{align}
\eta_0  &= -\left(1-\frac{\eta}{d+2}\right) 
\frac{\tilde g}{( \tilde m+1)^3}  \bigg[
\frac{27 \tilde m^2 \left(\frac{\tilde \lambda  \tilde m}{\tilde g^2}+1\right)^3}{2 (\tilde m+1)^2} \nonumber\\ 
&+\frac{9 \tilde m \left(1-\frac{11 \tilde \lambda  \tilde m}{4 \tilde g^2}\right) \left(\frac{\tilde \lambda  \tilde m}{\tilde g^2}+1\right)}{2 (\tilde m+1)}\left(1-\frac{3 \tilde \lambda  \tilde m}{2 \tilde g^2}
\right)
\bigg].
\end{align}
We note that the analysis presented in Ref.~\onlinecite{Chiocchetta2016} led to a wrong expression for \(\eta_0\), given by Eq.~(G6) (not reported here). In fact, as discussed here in the main text, they missed to add to \(r\) the term proportional to \(\lambda\) (see Eq.~\eqref{massn3} for the correct equation for \(r_k\)).
 The computation of \(\theta\) follows via Eq.~\eqref{thetashorttime}, once the dimensionless fixed-point values of \(m_k\), \(g_k\) and \(\lambda_k\) are calculated. For instance, using the values Eq.~\eqref{eq:FPvaluess}, one obtains \(\theta\) for \(d=3\). Finally the universal amplitude ratio \(X^\infty\) is calculated accordingly to Eq.~\eqref{Xlpa}.

\bibliography{biblio}
\end{document}